\documentclass[twocolumn, switch]{article}
\usepackage{preprint}
\usepackage{epsfig}
\usepackage{endnotes}
\usepackage{tikz}
\usepackage{amsmath}
\usepackage{graphicx}
\graphicspath{ {./images/} }
\usepackage{caption}
\usepackage{subcaption}
\usepackage{listings}
\usepackage{fontawesome}
\usepackage{svg}
\usepackage{comment}
\usepackage{xcolor}
\usepackage{xspace}
\usepackage{booktabs}
\usepackage{colortbl}
\usepackage{multirow}
\usepackage{graphicx}
\usepackage{lscape}
\usepackage{rotating}
\usepackage{float}
\usepackage{url}
\usepackage{cite}

\definecolor{codegreen}{rgb}{0,0.6,0}
\definecolor{codegray}{rgb}{0.5,0.5,0.5}
\definecolor{codepurple}{rgb}{0.58,0,0.82}
\definecolor{backcolour}{rgb}{0.95,0.95,0.92}

\newcommand{\ph}[1]{{\color{red}$<$}\textsc{PH}\xspace{\color{red}$>$} }
\newcommand{\thesystem}{\textsc{Lase}\xspace}
\newcommand{\ak}[1]{$<$ \textbf{\textit{AK: }}{\color{blue} #1} $>$}
\newcommand{\jz}[1]{$<$ \textbf{\textit{JZ: }}{\color{olive} #1} $>$}
\newcommand{\lr}[1]{$<$ \textbf{\textit{LR: }}{\color{codepurple} #1} $>$}

\lstdefinestyle{mystyle}{
    backgroundcolor=\color{backcolour},   
    commentstyle=\color{codegreen},
    keywordstyle=\color{magenta},
    numberstyle=\tiny\color{codegray},
    stringstyle=\color{codepurple},
    basicstyle=\ttfamily\footnotesize,
    breakatwhitespace=false,         
    breaklines=true,                 
    captionpos=b,                    
    keepspaces=true,                 
    numbers=left,                    
    numbersep=5pt,                  
    showspaces=false,                
    showstringspaces=false,
    showtabs=false,                  
    tabsize=2
}
\begin{document}

\title{An In-kernel Forensics Engine for Investigating Evasive Attacks}

\author{ Javad Zandi \\
	Florida International University\\
	\And
	Lalchandra Rampersaud \\
	Florida International University\\
 \And
	Amin Kharraz \\
	Florida International University\\
}

\twocolumn[\begin{@twocolumnfalse}

\maketitle

\begin{abstract}
Over the years, adversarial attempts against critical services have become more effective and sophisticated in launching 
low-profile attacks. 
This trend has always been concerning. However, an even more alarming trend is the increasing difficulty of collecting relevant evidence about these attacks and the involved threat actors in the early stages before significant damage is done.
This issue puts defenders at a significant disadvantage, as it becomes exceedingly difficult to understand the attack details and formulate an appropriate response.

Developing robust forensics tools to collect evidence about modern threats has never been easy. 
One main challenge is to provide a robust trade-off between achieving sufficient visibility while leaving minimal
detectable artifacts. This paper will introduce \thesystem, an open-source Low-Artifact Forensics Engine to perform threat analysis and forensics in Windows operating system. \thesystem augments current analysis tools by providing detailed, system-wide monitoring capabilities while minimizing detectable artifacts. We designed multiple deployment scenarios, showing \thesystem's potential in evidence gathering and threat reasoning in a real-world setting. By making \thesystem and its execution trace data available to the broader research community, this work encourages further exploration in the field by reducing the engineering costs for threat analysis and building a longitudinal behavioral analysis catalog for diverse security domains.

\end{abstract}
\vspace{0.5cm}

\end{@twocolumnfalse}]


\section{Introduction}
\label{Introduction}

Modern attacks against critical infrastructure                 
are continuously getting cheaper, faster, and more consequential.
For instance, in December 2021, just a few days after the Log4j vulnerability was disclosed~\cite{cisco}, adversaries started to adapt their code to locate exposed log4j services on the Internet and actively exploited this vulnerability. Very recently, the Clop ransomware group claimed responsibility for attacking the MOVEit transfer service on several known financial, education, U.S. federal, and state governments over the Memorial Day holiday. The attack was so consequential that the U.S. State Department announced a \$10 million bounty for information on the actors and involved threat campaign~\cite{state}. These incidents, on the one hand, suggest that adversaries are opportunistic and have asymmetric power in repurposing their tools to target new vulnerabilities in exposed services. However, a more concerning issue is that collecting necessary forensic evidence about these incidents at the early stages of the attacks before they cause consequential damage to the target systems is often challenging.

\noindent Over the years, the security community has invested significant effort to close this gap by developing solutions to extract insights from large volumes of raw network~\cite{fu2022encrypted, goyal2023sometimes, apruzzese2023sok} or default logging traces~\cite{kirat2014barecloud, goyal2023sometimes, hassan2020tactical, Hassaan2021IEEE, zipperle2022provenance, Yushan2018towards, gandhi2023rethinking} that primarily reflect temporal usage of systems resources (e.g., CPU, RAM, Network, Disk). 
While these efforts have been instrumental in detecting specific forms of attacks with predictable patterns (e.g., Ransomware~\cite{kharraz2015cutting}, crypto mining\cite{tahir2017mining}), they often fail to manifest fine-grained information on attack techniques that implement not evident or less known evasive techniques to deliver their payloads. Consider an attack where the offending process injects a payload into another process and uses that as a proxy to establish a backdoor to a remote server.
This incident, as a sequence of system-wide temporal data, is not collected effectively by engines that rely on default logging mechanisms.
This insufficient visibility over the dynamic behavior of malicious operations can put defenders in a highly disadvantaged
position to understand the attack tactics and lateral movements, perform root cause
analysis, and formulate a proper response. 

\noindent The core \textbf{insight} in this paper is today's threat intelligence against modern evasive attacks has to provide fine-grained system-wide visibility, leave minimal detectable artifacts, and be portable. These design goals have been historically difficult to attain at the same time. For instance, having fine-grained temporal visibility requires looking into very low-level components of the systems, which could make the solution less portable. On the other hand, approaches such as API hooking offer great portability, but there are several ways to bypass them. 
This paper aims to move towards addressing this need by introducing \thesystem -- an open-source \textbf{L}ow-\textbf{A}rtifact Foren\textbf{S}ics \textbf{E}ngine, that aims to define a balance 
between offering \emph{fine-grained visibility} while \emph{minimizing detectable} artifacts. These two design requirements are critical to staying effective over time in a landscape where adversaries actively try to identify defense solutions. To this end, \thesystem is deployed as a system-wide in-kernel engine that operates in high-privileged mode, making it almost impossible for user-mode applications to fingerprint, tamper, or kill the engine. At the same time, it offers system-wide visibility, temporal data on processes and threads, I/O requests, synchronous and asynchronous I/Os, fast I/Os, which are essentials to record the behavior of various forms of evasive attacks.

\noindent We highlight the portable design of \thesystem through two distinct security-motivated case studies: (1) an in-kernel baremetal-assisted threat analysis environment: we deployed the engine on several live-running physical machines to perform large-scale malware analysis on baremetal Windows, machines(Section~\ref{subsec:usecase1}), 
(2) a distributed deception-based infrastructure: We deployed the \thesystem-enabled images as in-cloud deception-based threat infrastructure for 46 days and collected attack artifacts about real-world attacks on intentionally vulnerable systems in the wild (Section~\ref{subsec:usecase3}). In the following, we summarize the key findings of the paper. 

\noindent The experiment shows that \thesystem-enabled baremetal analysis environment can complement current malicious code analysis tools on large-scale threat analysis and measurement.
We created Windows 10 OS images enabled by \thesystem and deployed them across eight physical machines, forming one of the first scalable baremetal-assisted analysis environments. We analyzed 79,544 malicious binaries over 30 months, from June 2022 to December 2024.  
Our analysis of recorded activities shows that a low-artifact engine can increase visibility over the behavior of malicious code. 
For instance, we collected 5,284,059 files after executing 34,438 malware samples that have file write activities in baremetal and virtualized environments. We observed that 3,981,555 (75.35\%) of the files were delivered only in the baremetal environment, and 884,301(16.74\%) were only dropped in the virtualized environment. This leaves 418,203 files that were delivered in both -- showing that the samples fetched different payloads. 
We have shared 2\% of the dataset (117GB) for this submission, which contains filesystem artifacts for bare-metal and virtualized analysis environments.

\noindent The analysis shows the emerging trend of using hardware-based fingerprinting makes malicious code analysis increasingly more challenging. In particular, among all forms of fingerprinting techniques we observed in the dataset, 1,187 of the samples were calling Windows Management Instrumentation (WMI) to detect advanced virtualized analysis systems (e.g., Hypervisor-based sandbox), 9,254 samples from Sabsik, Wacatac, AgentTesla families checked for direct CPU clock access, and 7,147 samples checking the BIOS hardware, firmware version, manufacturer, and configurations loaded during the system boot process as shown in table \ref{tbl:fingerprinting}. While advanced sandboxes may attempt to mask these artifacts and report different driver versions or hardware information, such modifications significantly reduce the robustness of the target system. That is,  these changes may result in a complete system crash or disruption of legitimate drivers because critical services often need exact information of hardware information for critical operations (e.g., integrity checking, patch and update, driver management). 

\noindent Finally, \thesystem was deployed as a portable forensics engine in a deception-based threat intelligence environment for 46 days by intentionally exposing vulnerable Windows services on the \thesystem-enabled cloud hosts. During the course of the experiment, we identified 734 successful intrusions. Our analysis shows that it takes approximately 6 hours for adversaries to discover the exposed vulnerable service and 13 hours to compromise the service. The analysis of the process and filesystem traces shows the execution of 1,221 executables, a collection of 401 shell scripts, 235 installers, 14,706 source code files, 438 digital certificates, and 4,586 custom dynamic libraries.

\noindent The proposed forensics engine and discussed use cases aimed to highlight a critical gap in security defense for portable yet robust forensics tools for attack analysis, bringing more behavioral visibility wihout introducing trivial detectable artifacts.
Our hope is that this work serves to raise awareness about the importance of forensics analysis frameworks for global visibility and intelligence gathering in today's attack landscape. We also hope that our approach will prove helpful to the security community in identifying emerging threats that go beyond what is routinely observed today and open the door to future research on threat intelligence and malicious code analysis. 

\noindent \textbf{Contributions.} The paper makes the following contributions:

\begin{itemize}
    \item We propose an in-kernel open-source forensics engine that can be deployed in all modern versions of Windows operating systems. The forensics engine collects almost all forensically relevant information at the process and thread level as well as I/O activities for analyzing run-time behavior of malicious.

    \item We deployed the engine in two case studies, showing the benefits of a low overhead 
    security forensics engine in providing insights into the threat landscape. We collected artifacts about 734 successful attacks on our deception environment. We also analyzed over 79K samples from hundreds of malware families and collected over 2.6 TBs of compressed threat artifacts over 30 months of experiments. 

    \item The source-code\footnote{https://tinyurl.com/LASECode} as well as the artifacts used in the analysis of the paper (2\% of the data catalog)\footnote{https://tinyurl.com/LASEArt} are made available. The distribution of malware families for the artifacts submitted can be found in Table \ref{tab:malware_distribution_in_artifacts}.
  
\end{itemize}

\section{Background / Motivation}
\label{sec:Background}

\noindent In this section, we begin by describing the threat model to describe attackers' capabilities in designing and developing malicious code and evasion mechanisms employed to bypass possible defense solutions. We then motivate the work by describing what is lacking in the current solutions to bring more visibility into the attack landscape.



\subsection{Threat Model}

\noindent In this paper, we assume that adversaries have significant freedom to develop evasive malicious code, as documented in prior work and community reports. In particular, our assumptions about attackers’ capabilities are as follows:

\noindent \textbf{Environment Sensitive:} There is no lack of evidence that modern evasive attacks are armed with various forms of anti-analysis techniques~\cite{kirat2014barecloud, miramirkhani2017spotless, Afianian2020survey, Bulazel2017survey, chen2016advanced, d2020dissection, kolbitsch2011power, chen2008towards, sikorski2012practical, polino2017measuring, lindorfer2011detecting}.
A long list of those artifacts (e.g., specific processes, registry keys, services, and network adapters) is publicly reported~\cite{MostcommonFingerprintingMethods}. These techniques are often used by malicious code developers to successfully intrude on a system where they plan to launch code.

\noindent \textbf{Debugging Resistant:}
Adversaries have historically incorporated anti-debugging techniques to complicate forensic analysis. 
In particular, one common debugging approach relies on code injection techniques~\cite{kuechlerndss21, alzahrani2022analysis, willems2007toward, brubacher1999detours} in the context of the target payload. 
That is, analysis agents (i.e., DLL) are injected into the context of a malicious process for behavioral monitoring. There have been several techniques to automatically detect such analysis techniques by simply listing active processes on target systems to identify debuggers or creating snapshots of their own memory structures, such as their heap and modules (DLLs loaded in the process’s virtual address space), to identify if they are being analyzed. Adversaries can incorporate anti-debugging techniques at different layers, including CPU registers, in-memory data structure checks, or calling specific APIs or native functions~\cite{plumerault2021dbi, willems2007toward, brubacher1999detours, galal2016behavior, marhusin2008evaluation, lopez2017survey}.

\noindent \textbf{Launching Proxy Operations:} Malicious processes often inject the actual malicious code into the context of legitimate processes to bypass reputation-based services~\cite{monnappa2017detecting, mohanta2020code}. This is an important assumption as it makes most debugging techniques that target particular processes ineffective. Hence, our threat model in this project will consider these and other similar risks, and the solutions that will be developed should be robust and tested against these malicious activities~\cite{hosseini2018ten, Jullian2018indepth, mohanta2020code, monnappa2017detecting}.

\noindent In this study, we also assume that user-initiated processes solely operate in user mode. Consequently, interactions with lower-level system resources are channeled through operating system API calls, which can be intercepted by a low-level forensic engine operating in the kernel space.
In addition, we also assume that the trusted computing base
includes the OS kernel and underlying software and hardware
stack, and that 
normal user-based access control prevents attackers from running malicious code with superuser privileges. 

\subsection{The Need for a Low-Artifact Forensics Engine}
\label{subsec:The Need for a Low-Artifact Forensics Engine}
\noindent To respond effectively to highly evasive methods described in the threat model, the security community has innovated at various layers to define their analysis frameworks\cite{kharaz2016unveil, kirat2011barebox, kirat2014barecloud, hassan2020tactical, erin2021usenixsec, goyal2023sometimes}, depending either on whether they found it the most effective way or where it was easier to incorporate the proposed method. In this section, we describe the limitations of those methods and make the case for how the forensics engine should be implemented to analyze the majority of threats.

\noindent \textbf{User-mode Hooking Techniques.} Classic hooking techniques~\cite{lopez2017survey, marhusin2008evaluation, brubacher1999detours,hosseini2018ten, shaid2015memory} often rely on inline overwriting of APIs which requires
injecting custom-formulated payloads into the context of the target process address space to identify the invoked functions, which inherently leaves several artifacts (e.g., checking the integrity
of the API function, monitoring the list of loaded processes). 
Furthermore, this is not a scalable approach due to the engineering efforts required to overwrite the target APIs. These modifications often are not without problems and could cause serious issues (e.g., system crashes) because of unexpected exceptions and unhandled cases. Lastly, function hooking or debugging mechanisms are inherently designed to be single process-centric and often lose their effectiveness rapidly when a more system-wide behavioral analysis is
required. Furthermore, subverting these mechanisms is a common practice among adversaries by copying the desired malicious code into the address space of another process or using other common hooking evasion techniques such as customized code, stole code~\cite{kawakoya2019api, kawakoya2013api, cheng2021obfuscation, starinkunderstanding, korczynski2017capturing}, and sliding calls~\cite{kawakoya2019api}.

\noindent \textbf{Built-in Logging and Diagnosis Platforms.}
Modern operating systems are often equipped with integrated logging platforms designed for event monitoring and application performance tracking (e.g., syslog)~\cite{essay83142}. In Windows, Event Tracing for Windows (ETW)~\cite{ETW-sec} serves as a logging tool for system-wide monitoring and performance data collection. 
While ETW trace recording has been useful for collecting security events (e.g., authentication logs), it was not primarily designed to stay undetected. Adversaries can enumerate active ETW providers that use it for analysis/inferencing purposes to list defense tools on target machines. This technique is often used to detect anti-malware tools on the target system. Furthermore, a user in an active session can disrupt the logging process by temporarily modifying ETW environment variables to redirect logs, disable logging in a specific context, or modify behaviors in an ETW-reliant application. That said, ETW offers significant visibility over the overall dynamic behavior of systems. However, the approach is vulnerable to fingerprinting techniques and disruption~\cite{blackhat-etw}. In Table~\ref{tab:list_of_edrs}, we also discuss a set of open-source forensics engine projects that are built on top of ETW service and compare with \thesystem. 







\noindent \textbf{Hypervisor-based Approaches.} Hypervisor-based methods~\cite{garfinkel2007compatibility, franklin2008remote, branco2012scientific} have been proposed to address the above shortcomings by putting the analysis module out of the operating system to achieve increased visibility and analysis capabilities (e.g., memory extraction, low-cost analysis, and debugging). 
While these techniques offer significant flexibility for code analysis, there are still low-cost techniques to fingerprint hypervisor-based solutions by, for instance, abusing critical Windows core APIs (e.g., Windows Management Instrumentation~\cite{aboutwmi, tanana2020behavior}) to extract hardware information or direct access to the CPU for detecting instruction execution delay. Disabling or overwriting these APIs is possible. However, these changes can be consequential because the normal operation of the many critical software components depends on accurately extracting hardware system information. Consequently, modifying these APIs can result in frequent system crashes and legitimate service disruptions.

\subsection{Design Requirements}
\label{sec:design-requirement}
\noindent Given the benefits and shortcomings of current methods, we believe that an in-kernel behavioral recording method is an appropriate place to collect forensic data for a wide range of applications in threat analysis. In the following, we describe the security requirements we are seeking to achieve acceptable visibility without leaving detectable artifacts.

\noindent \textbf{Low Artifact Operations.} 
Low-artifact behavioral monitoring is a critical property in analyzing modern threats (e.g., analyzing malicious code, recording techniques, tactics, and procedures) because, in almost all of these incidents, adversaries' goal is to identify signs of an analysis environment. As mentioned earlier, approaches such as user-mode hooking techniques or hypervisor techniques introduce specific forms of artifacts that can be weaponized by adversaries to analyze the environment. New threat intelligence frameworks should make the target systems more robust against these fingerprinting efforts.

\noindent \textbf{System-wide Behavioral Monitoring.}
System-wide behavioral monitoring refers to the property that enables run-time behavioral monitoring in a multi-process environment. This is a critical property in modern forensic analysis towards achieving complete mediation since malicious code can launch the actual malicious payload via proxy operations (as mentioned in the threat modeling). A system-wide view can provide sufficient visibility about the interaction of a process not only with the operating system resources but also with other processes~\cite{cowan2006turing}. From a threat intelligence standpoint, system-wide monitoring helps to establish a clear and comprehensive audit trail of all actions taken on a system by users or processes. 

\noindent \textbf{OS Support.} 
It is critical that monitoring mechanisms operate in a high-privileged mode 
while introducing low integration costs. To this end, in-kernel forensics layer techniques are viable design choices to achieve these goals due to the significant OS support. An in-kernel forensics engine inherently runs as a privileged operation in the kernel-level~\cite{gupta2022popkorn}. Therefore, termination or manipulation of core functionalities will not be possible by the malicious process
that operates at the user level. In particular, Windows officially supports the concept of protected services, called Early Launch Anti-Malware (ELAM), to allow critical services to run as protected services~\cite{korkin2021protected}. After the service is launched, Windows incorporates a code integrity mechanism to only allow trusted code to load into a protected service. Windows also protects these services from code injection~\cite{milenkoski2019elam, OverELAM}. This approach will guarantee that the forensics engine is protected from common forms of availability, memory corruption, tampering, and code injection attacks.

\noindent Considering all design options and the proposed threat model, an in-kernel behavioral recording method appears to be a suitable layer for enabling a portable and low-artifact forensics layer to study common threats in the wild.
That is, the in-kernel module does not modify any Windows APIs or inject any code into running processes. Furthermore, the flexibility to deploy the engine in baremetal systems makes the entire analysis system robust against advanced fingerprinting techniques that target hardware specifications. \emph{That said, this design should not be considered as an alternative solution to the current hypervisor-based solution that can offer significant reverse engineering flexibility. Rather, it can serve as the first line of defense in threat analysis by satisfying
fingerprinting checks that are impossible or costly in other defense systems.}

\section{Monitoring Run-Time Behavior}
\label{sec:methodology}

\begin{figure}[htb!]
    \centering
    \includegraphics[width=0.95\linewidth]{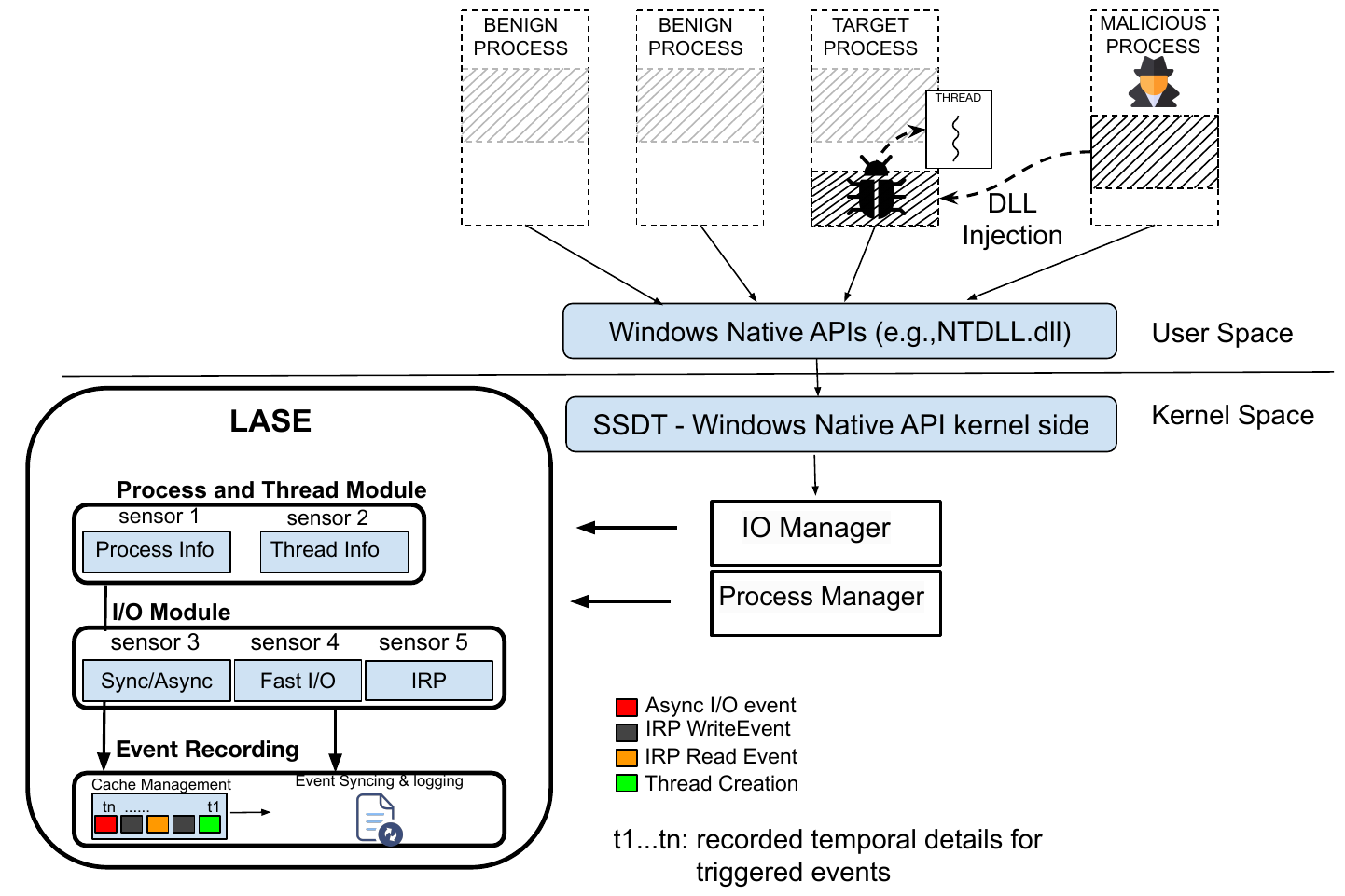}
    \caption{The high-level view of \thesystem's architecture. \thesystem provides a system-wide multi-process forensics engine by defining three in-kernel modules: (1) Process and thread module, (2) Filesystem and I/O module, (3) Event recording module. In this figure, \thesystem records run-time behavior of the malicious process that uses DLL injection to launch a proxy attack.} 
    \label{fig:lase_system_architecture}
\end{figure}


\noindent In this section, we discuss the engineering decision and details of \underline{L}ow-\underline{A}rtifact Foren\underline{S}ics \underline{E}ngine (\thesystem). We elaborate on how we implemented the forensics engine. We provide examples of the collected artifacts and the deployment details of \thesystem to better analyze modern threats. 

\subsection{Forensics Agents and Major Components} This section describes the main modules of \thesystem that enables run-time trace recording.  

\noindent \textbf{I. Process and Thread Monitoring.} Modern malicious code often employs code injection techniques such as DLL injection or process hollowing to inject code into legitimate processes~\cite{kawakoya2019api, ray2012defining, mitropoulos2017fatal, barabosch2014host, barabosch2017quincy, cheng2018towards, korczynski2017capturing}. This allows the malicious code to execute in the context of a legitimate process and and run proxy attacks. Detecting such evasive operations has never been trivial since adversaries have significant freedom on what process to choose for injection, and making any assumption about the target process can impact the visibility factor. To this end, \thesystem records system-wide monitoring for processes and threads 
by recording process-related events such as \emph{initiation and termination}, thread-level events such as \emph{creation and deletion}, and process image loading events such as \emph{load and unload}.
This driver takes advantage of the Windows kernel APIs located in the \texttt{NtosKrnl.lib}~\cite{ntddk, hai2023proposed, jones2012ransomware} library and invokes embedded Windows Kernel functions: \texttt{PsSetCreateProcessNotifyRoutineEx()}~\cite{PsSetCreateProcessNotifyRoutineEx}, \texttt{PsSetCreateThreadNotifyRoutineEx()}~\cite{PsSetCreateThreadNotifyRoutineEx} and \texttt{PsSetLoadImage\\NotifyRoutineEx()}~\cite{PsSetLoadImageNotifyRoutineEx} for process, thread and image activities respectively. Each of these functions allows the \thesystem driver to register a callback routine, which is registered in the kernel data structure within the process management subsystem and called by the \texttt{Process Manager}~\cite{yosifovich2017windows} whenever its corresponding event occurs. These functions are part of the process and thread management in Windows and provide mechanisms for kernel-mode components to monitor process lifecycle events.

\noindent \textbf{II. I/O Request Packets.}
The execution of input and output operations in the Windows operating system, specifically related to driver operations, is facilitated by the utilization of packets. These packets, known as I/O packets or IRPs (I/O Request Packets), include encoded information that directs and controls the driver's actions~\cite{probert2010windows, russinovich2012windows}. The self-contained structure of the Input/Output Request Packet (IRP) encompasses all the necessary information for a driver to manage an I/O request. To enhance operational efficiency and contextual relevance, IRPs are classified into major and minor operations. The major function code instructs the driver on what action to take to fulfill the I/O request. Minor codes refer to a specific category of major operations and supply further detailed information to the driver. A major function code specifies the specific type of I/O operation in conjunction with an optional minor function code~\cite{probert2010windows, WPnPIRPMinor}. Recording the forms of IRPs shown in table~\ref{tab:irp descriptions} in the Appendix is critical to understanding different classes of system interactions. That is, all filesystem activities, such as fetching files into the memory or interacting with cached files, are translated into a set of IRP requests that can be captured in the kernel. \thesystem monitors all 103 IRPs, including 45 major and 48 minor I/O request packets (IRPs) based operations encompassing persistent and cached files.


\noindent \textbf{III. Synchronous and Asynchronous I/O.}
In Windows, most of the I/O operations generated by processes are synchronous~\cite{sync}. That is, a process waits until the OS returns a status code when the I/O is complete. For instance, ReadFile and WriteFile functions are executed synchronously. However, programs can modify the default behavior and issue asynchronous I/O requests by generating multiple I/O requests and continue executing while the OS performs the I/O operations. Our analysis shows that asynchronous behavior is largely ignored by almost all the current solutions~\cite{kirat2014barecloud, kharaz2016unveil, spensky2016phi, kirat2011barebox}. This is an important property because it allows processes to launch concurrent malicious operations, evading defense solutions that wait for the immediate return of the status code.



\noindent \textbf{IV. Fast I/O.} Programs can be designed to use Fast I/O requests for reading and writing to cached files, enabling rapid access synchronously~\cite{fastio}. In Fast I/O operations, the data is transferred directly between user buffers and the cache. This operation can bypass all current mechanisms that generate an IRP. For instance, if a process generates a Fast I/O operation requesting access to the content of an opened file in the memory, the requests will be processed immediately without generating any IRP events. This is an important observation since adversaries or malicious code can generate Fast I/O requests to access open files in the system buffer without generating any IRP artifacts~\cite{solomon2009windows}.

\noindent \textbf{V. Event Recording.} 
The event recording module operates as a producer-consumer architecture~\cite{byrd1999producer} where producers generate filesystem and process events, and consumers use the events for processing, which include storing on disk and sending high-priority events to the network.
We developed a trace recording library that offers a set of Application Programming Interfaces (APIs) to store the events and manage I/O event monitoring. The library provides a mechanism to create data containers to maintain I/O traces in the memory for further processing. Generating the data containers requires close interaction
with the OS cache manager to allocate the necessary memory space.
One approach to implement the in-memory data cache is a write-back cache~\cite{ghandeharizadeh2019design} where
the data is updated only in the cache, and permanent storage is updated later. The write-back cache is a straightforward caching mechanism and is an excellent fit for general read and write I/O events. However, continuous I/O monitoring can significantly increase the required cache size on machines with limited memory and storage capacity. As an alternative approach, we implement a circular buffer mechanism using a queue, where I/O events are broken into smaller and more manageable data chunks for processing and consumption. The caching policy is implemented as a multi-threading model to deliver data from multiple producers to multiple consumers concurrently.

\subsection{I/O Benchmarks}
\noindent \thesystem operates as an in-kernel module to record several forensics events. Consequently, it is important to evaluate the imposed overhead in a quantifiable fashion. To this end, we 
conducted an experiment to evaluate the overhead on filesystem operations as they are the more common and intense operations due to the customized instrumentation layer and log collection process. 
We generated workloads by creating 500 small (10Kb) and large (10Mb) files to test the throughput of block
write, rewrite, and read operations. We used these operations as the major events that require significant filesystem interaction. 
The disk I/O performance was assessed using the popular Windows file system benchmarking tool, IOZone~\cite{iozone} by measuring the time to process the creation and access of generated files. Each experiment was repeated ten times for 4.42 minutes, and the average score was calculated to yield the final results presented in table \ref{tbl:lasebenchmark}. The experiments show that \thesystem performs well with both small and large files and imposes an overhead between 1.74\% and 5.31\%. We consider the 2.01\% overhead for reading an existing small file for the first time. %
One reason for the relatively lower overhead for these critical operations is the design decision to monitor forensics events with minimal changes to the standard subsystems. Note that \thesystem leverages an event-driven approach and does not require injecting code into the context of target processes. This approach has two main benefits: (1) it is a passive approach and avoids intrusive operations that may require any modification to the filesystem or standard OS functions, introducing lower I/O overhead compared to inline API modifications, (2) it minimizes possible interference with the normal operation of processes and
reduces the risk of crashes and memory corruption.

\begin{table}[!ht]
    \centering
    \caption{Disk I/O performance in a standard and host running \thesystem}
    \footnotesize
    \begin{tabular}{lrrr}
    \toprule
        \textbf{Operation}  & \textbf{Standard}  & \textbf{\thesystem} & \textbf{Overhead} \\ \midrule

\textbf{Writer}  & ~ & ~ & ~\\
         \hspace{3mm}Small & 788,641 KB/s & 808,474 KB/s & 2.51\%  \\ 
         \hspace{3mm}Large & 893,921 KB/s &  909,448 KB/s & 1.74\%  \\ 

        \textbf{Re-writer}  & ~ & ~ & ~\\
         \hspace{3mm}Small & 1,036,665 KB/s &   1,05,9468 KB/s & 2.20\% \\ 
         \hspace{3mm}Large & 1,054,297 KB/s &  1,092,756 KB/s & 3.65\% \\  

         \textbf{Reader}  & ~ & ~ & ~\\
         \hspace{3mm}Small & 3,564,507 KB/s & 3,492,892 KB/s& 2.01\%  \\ 
         \hspace{3mm}Large & 2,841,287 KB/s &  2,928,931 KB/s & 3.08\% \\ 

         \textbf{Re-Reader}  & ~ & ~ & ~\\
         \hspace{3mm}Small & 4,228,550 KB/s &  4,452,886 KB/s & 5.31\% \\ 
         \hspace{3mm}Large & 3,643,169 KB/s &  3,791,870 KB/s & 4.08\%  \\ \bottomrule
      
    \end{tabular}
    \label{tbl:lasebenchmark}
\end{table}



\begin{table*}[!ht]
\centering
\caption{Comparing \thesystem with current forensics engines}
\label{tab:list_of_edrs}
\resizebox{\textwidth}{!}{%
\begin{tabular}{lccccccc}
\toprule
\multicolumn{1}{c}{Recorded Activities} & LASE & BestEDR\cite{edr_BEOTM} & Fibratus\cite{edr_Fibratus} & Osquery\cite{edr_osquery} & Wazuh\cite{edr_wazuh} & BlueSpawn\cite{edr_BlueSpawn} & Whids\cite{edr_Whids} \\ \hline
\textbf{Technology} & \textbf{kernel driver} & API Hooking & ETW & App log collection & App log collection & Win32 API & ETW, Sysinternal \\ \hline
\textbf{Run-time Behavior} &  &  &  &  &  &  &  \\ \hline
Process-level Information & \textbf{system-wide} & single proc. & system-wide & single proc. & single proc. & single proc. & system-wide \\ 
Thread-level Information & \textbf{system-wide} & no thread & system-wide & no thread & no thread & single proc. & system-wide \\ 
DLL Injection & \textbf{system-wide} & single proc. & not supported & not supported &not supported & not supported& not supported \\
Inter-process Communication & \textbf{system-wide} & not supported & not supported & not supported & not supported& not supported& not supported\\ \hline
\textbf{I/O Activities} &  &  &  &  &  &  &  \\ \hline
Filesystem I/O & \textbf{system-wide} & not supported& system-wide & not supported & not supported& single proc. & system-wide \\
Synchronous I/O & \textbf{system-wide} & not supported & system-wide &not supported& not supported & single proc. & system-wide \\
Asynchronous I/O & \textbf{system-wide} & not supported & not supported&not supported & not supported& not supported& not supported \\
Fast I/O & \textbf{system-wide} & not supported& not supported & not supported& not supported & not supported & not supported \\
\textbf{Forensics Agents} &  &  &  &  &  &  &  \\ \hline
Logging Service & \textbf{kernelmode} & usermode & usermode & usermode & usermode & usermode & usermode \\
Scheduler Agent & \textbf{kernelmode} & usermode & usermode & usermode & usermode & usermode & usermode \\ \bottomrule
\end{tabular}%
}
\end{table*}

\subsection{Resistance Against Common Dynamic Evasion Mechanisms}       
\noindent 
In the following, we briefly describe common fingerprinting checks used by adversaries in the wild and discuss the robustness of \thesystem against each method.

\noindent \textbf{VM Checks.} Modern attacks are sensitive to virtualized environments and may not run correctly in those systems~\cite{erin2021usenixsec}. Checking registry keys (e.g., reg\_key, reg\_key\_value)~\cite{179470}, firmware details (e.g., firmware ACPI, RSMB), CPU information (e.g., cpuid)~\cite{179470} and hardware information (e.g., model\_computer\_system\_wmi) are just a few examples of such fingerprinting techniques~\cite{miramirkhani2017spotless}. \thesystem is portable and can be deployed on baremetal machines (see case study 1). Consequently, none of the VM checks would apply to the threat infrastructure enabled by \thesystem. We should also highlight that \thesystem-enabled infrastructure is robust against advanced fingerprinting techniques (e.g., device information retrieval through Windows Management Instrumentation)~\cite{wmi-2} used to fingerprint hypervisor-based environments given that the reference platform does not use any virtualization technologies vulnerable to such fingerprinting attempts.

\noindent \textbf{Anti-Debugging Checks.} 
Anti-debugging techniques are strategies to thwart efforts to understand and dissect malicious software. Anti-debugging aims to make debugging more difficult, time-consuming, or outright impossible. For instance, it is quite common in malicious payloads to check for debuggers in various levels (e.g., SharedUserData\_KernelDebugger) by checking ports or debugging objects (e.g., processDebugPort, processDebugObject), interrupts, and hardware breakpoints(e.g., Interrupt\_0$\times$2d, SystemKernelDebuggerInformation)~\cite{guibernau2020catch,liu2022enhancing,galloro2022systematical}. \thesystem is an event-driven engine that relies on filter driver design to achieve its design goals.
It does not require injecting the debugging agent into the context of target processes to record activities, leaving almost no traces for anti-debugging checks initiated by the malicious process.

\noindent \textbf{Resource Profiling Checks.} Resource profiling checks complement anti-debugging and VM checks~\cite{liu2022enhancing}. For instance, checking the Windows data structure for processes (e.g., process\_enum)~\cite{179470,miramirkhani2017spotless}, or information on Disk (disk\_size\_getdiskfreespace, \\ disk\_size\_wmi)~\cite{wmi-2} allows adversaries to identify sandboxes or virtual environments~\cite{179470,galloro2022systematical}. Since \thesystem is deployed in a baremetal environment, these checks will not likely assist adversaries.


\subsection{Comparison with Other Open-Source Forensics Tools.}

\noindent As the last comparison, we analyzed publicly available forensics tools by comparing their architectural details, recorded artifacts, and fingerprintable details that could impact the effectiveness of these engines.

\noindent \textbf{Selection Metrics.} We evaluated forensics engines using three key criteria for a head-to-head comparison: the engine (1) must be open-source, (2) must be compatible with Windows operating systems, and (3) must be installed and operate without compilation or fundamental deployment errors. Our initial analysis included 10 engines, from which we selected six for this evaluation following these metrics.

\noindent \textbf{Analysis.} Table~\ref{tab:list_of_edrs} illustrates the summary of the analysis. We observed that user-mode API hooking is still a common practice. For instance, BestEDR~\cite{edr_BEOTM}, OSquery~\cite{edr_osquery}, Wazuh~\cite{edr_wazuh}, BlueSpawn\cite{edr_BlueSpawn} are all developed on top of API hooking or standard application logging mechanisms. Implementing this design strategy results in easier evasion and tampering since the forensic engine leaves detectable traces. Furthermore, tampering or terminating the engine is quite straightforward when the artifact recording engine operates at the same privilege level as other processes. Whids~\cite{edr_Whids} approximates \thesystem in terms of gaining visibility by augmenting Microsoft ETW. However, the log processing and the scheduling module reside in the user space, making the tool vulnerable to evasion and fingerprinting. 
As previously noted, ETW was initially developed for performance monitoring, and minimizing artifacts was not a primary design consideration. Furthermore, it is not designed to capture specific run-time security events, such as remote code injection, that can have significant security consequences. That said, \thesystem offers system-wide visibility over processes, threads, and I/O activities while maintaining all functionalities (e.g., logging, monitoring tasks) within the kernel. This can potentially make \thesystem a more generalizable solution for the same problem space. 

\subsection{Collecting Run-time Artifacts}
\noindent In this section, we discuss the format and details of the recorded artifacts. To this end, we provide an example of a macro-based malware sample where the malicious code loads and compiles visual basic code. Table \ref{tab: macro-based malware trace} depicts the low-level temporal artifacts collected by \thesystem in this attack. 
The sequence of commands provided illustrates a multi-faceted approach used by the macro-based malware to infiltrate a system, execute malicious payloads, and maintain persistence. In particular,
the malicious code uses Dynamic Data Exchange (DDE) to pass commands to Excel, a method known to execute code within an Excel document. The subsequent command (line 2) indicates the use of embedded mode, likely to run a macro embedded within the Excel document.

\begin{figure}[!ht]
\centering
\includegraphics[width=\columnwidth]{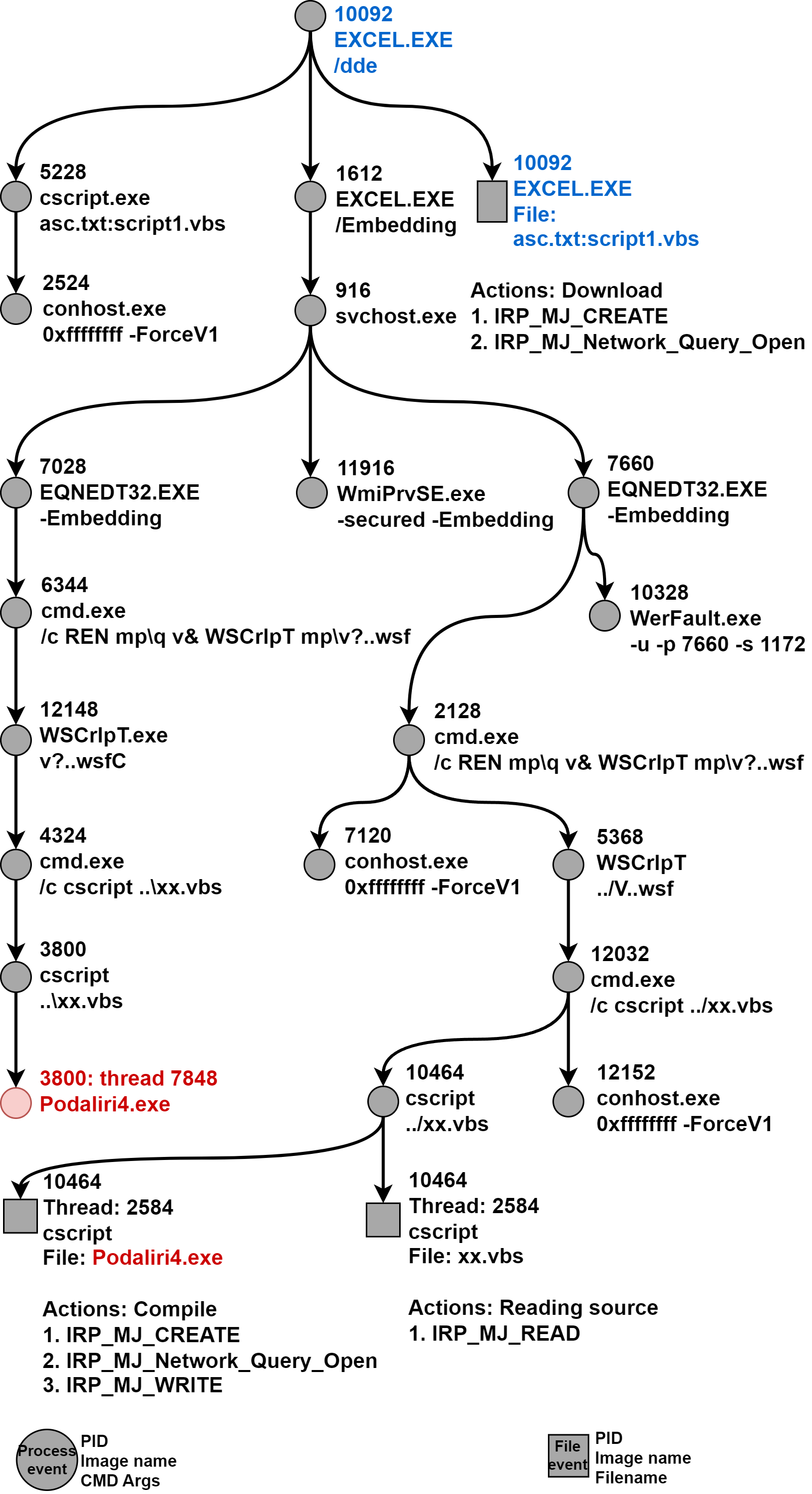}
\caption{The collected artifacts for a micro-based malware attack. The malicious process invokes 15 processes in the background to launch a successful attack.}
\label{fig:use_case_macro_malware}
\end{figure}

\noindent The repeated use of `eqnedt32.exe -Embedding' in the trace suggests the exploitation of vulnerabilities in the Equation Editor. Older versions of EQNEDT32.EXE have known security flaws that can be exploited to execute arbitrary code~\cite{gu2022minsib, elder2024automatic}.
Lines 14 to 25 show that the malware incorporates Windows command prompt and script hosts such as wscript.exe and cscript.exe to execute scripts. The filesystem activity shows that these scripts were used to download additional payloads (i.e., xx.vbs, Podaliri4.exe), modify system settings, and establish persistence. Using commands such as `WmiPrvSE.exe -secured -Embedding' invokes Windows Management Instrumentation (WMI) to perform system management tasks, including creating persistence and gathering system information. The artifact also shows that the malware uses Alternate Data Streams (ADS), a technique to hide malicious scripts within legitimate files, making them harder to detect. This multi-stage attack resulted in the launch of the final process, i.e., Podaliri4.exe, which delivered ransomware to the system. 
Figure \ref{fig:use_case_macro_malware} depicts the attack tree generated from the data collected.

\begin{table*}[!ht]
\caption{\thesystem artifacts for a multi-stage macro-malware attack (md5:7ccf88c0bbe3b29bf19d877c4596a8d4). The malware incorporates several techniques for fetching malicious payload, information grabbing, persistence, and camouflage.}
\label{tab: macro-based malware trace}
\centering
\begin{subtable}{0.9\textwidth}
\centering
\resizebox{\textwidth}{!}{ 
\begin{tabular}{ccccccclll}
\hline
\multicolumn{1}{|c|}{\#} & \multicolumn{1}{|c|}{Operation} & \multicolumn{1}{c|}{Time}  & \multicolumn{1}{c|}{Global \#}  & \multicolumn{1}{c|}{PPID} & \multicolumn{1}{c|}{PID}  & \multicolumn{1}{c|}{TID}  & \multicolumn{1}{c|}{Image path}  & \multicolumn{1}{c|}{Command-line args}  & \multicolumn{1}{c|}{File path}  \\ \hline
\rowcolor[HTML]{CBCEFB} 
1  &  Pr Create  & 20:51:45:628  & 183668  & 5480  & 10092  & 0  & \%MSOffice\%\textbackslash{}EXCEL.EXE  & /dde  &  \\
\rowcolor[HTML]{CBCEFB} 
2  &  Pr Create  & 20:51:48:141  & 205076  & 10092  & 1612  & 0  & \%MSOffice\%\textbackslash{}EXCEL.EXE  & /Embedding  &  \\
\rowcolor[HTML]{CBCEFB} 
3  &  Pr Create  & 20:52:16:301  & 346364  & 916  & 7028  & 0  & \%MSOfficeCommon\%\textbackslash{}eqnedt32.exe & -Embedding  &  \\
\rowcolor[HTML]{CBCEFB} 
4  &  Pr Create  & 20:52:16:579  & 348340  & 7028  & 6344  & 0  & \%SysWOW64\%\textbackslash{}cmd.exe  & /c REN mp\textbackslash{}q v\& WSCrIpT mp\textbackslash{}v?..wsf C &  \\
\rowcolor[HTML]{CBCEFB} 
5  &  Pr Create  & 20:52:16:668  & 350587  & 6344  & 12148  & 0  & \%SysWOW64\%\textbackslash{}wscript.exe  & \%TEMP\%\textbackslash{}v?..wsf C  &  \\
\rowcolor[HTML]{CBCEFB} 
6  &  Pr Create  & 20:52:17:084  & 358524  & 12148  & 4324  & 0  & \%SysWOW64\%\textbackslash{}cmd.exe  & /c cscript \%TEMP\%\textbackslash{}xx.vbs  &  \\
\rowcolor[HTML]{CBCEFB} 
7  &  Pr Create  & 20:52:17:149  & 359955  & 4324  & 3800  & 0  & \%SysWOW64\%\textbackslash{}cscript.exe  & \%Temp\%\textbackslash{}xx.vbs  &  \\
\rowcolor[HTML]{FFFFC7} 
8  &  Pr Exit  & 20:52:17:520  & 376991  & 916  & 7028  & 0  & \%MSOfficeCommon\%\textbackslash{}eqnedt32.exe &  &  \\
\rowcolor[HTML]{FFFFC7} 
9  &  Pr Exit  & 20:52:18:638  & 380165  & 10092  & 1612  & 0  & \%MSOffice\%\textbackslash{}EXCEL.EXE  &  &  \\
\rowcolor[HTML]{CBCEFB} 
10  &  Pr Create  & 20:52:19:456  & 381227  & 916  & 11916  & 0  & \%SysWOW64\%\textbackslash{}wbem\textbackslash{}WmiPrvSE.exe & -secured -Embedding  &  \\
\rowcolor[HTML]{CBCEFB} 
11  &  Pr Create  & 20:53:05:751  & 407732  & 10092  & 5228  & 0  & \%SysWOW64\%\textbackslash{}cscript.exe  & C:\textbackslash{}programdata\textbackslash{}asc.txt:script1.vbs &  \\
\rowcolor[HTML]{FFFFC7} 
12  &  Pr Exit  & 20:57:10:385  & 589717  & 5480  & 10092  & 0  & \%MSOffice\%\textbackslash{}EXCEL.EXE  &  &  \\
\rowcolor[HTML]{CBCEFB} 
13  &  Pr Create  & 20:57:42:131  & 666687  & 916  & 7660  & 0  & \%MSOfficeCommon\%\textbackslash{}eqnedt32.exe & -Embedding  &  \\
\rowcolor[HTML]{CBCEFB} 
14  &  Pr Create  & 20:57:42:305  & 668655  & 7660  & 2128  & 0  & \%SysWOW64\%\textbackslash{}cmd.exe  & /c REN mp\textbackslash{}q v\& WSCrIpTmp\textbackslash{}v?..wsf  C &  \\
\rowcolor[HTML]{CBCEFB} 
15  &  Pr Create  & 20:57:42:358  & 671600  & 2128  & 5368  & 0  & \%SysWOW64\%\textbackslash{}wscript.exe  & \%Temp\%\textbackslash{}v?..wsf C  &  \\
\rowcolor[HTML]{CBCEFB} 
16  &  Pr Create  & 20:57:42:399  & 675850  & 7660  & 10328  & 0  & \%SysWOW64\%\textbackslash{}WerFault.exe  & -u -p 7660 -s 1172  &  \\
\rowcolor[HTML]{CBCEFB} 
17  &  Pr Create  & 20:57:42:474  & 678590  & 5368  & 12032  & 0  & \%SysWOW64\%\textbackslash{}cmd.exe  & /c cscript \%Temp\%\textbackslash{}xx.vbs  &  \\
\rowcolor[HTML]{CBCEFB} 
18  &  Pr Create  & 20:57:42:516  & 679047  & 12032  & 10464  & 0  & \%SysWOW64\%\textbackslash{}cscript.exe  & \%Temp\%\textbackslash{}xx.vbs  &  \\
\rowcolor[HTML]{FFCCC9} 
19  &  Ld Image  & 20:57:42:527  & 679237  & 12032  & 10464  & 0  & \%SysWOW64\%\textbackslash{}cscript.exe  &  & \%SysWOW64\%\textbackslash{}bcryptprimitives.dll \\
\rowcolor[HTML]{FFCCC9} 
20  &  Ld Image  & 20:57:43:645  & 705589  & 12032  & 10464  & 0  & \%SysWOW64\%\textbackslash{}cscript.exe  &  & \%SysWOW64\%\textbackslash{}winhttp.dll  \\
\rowcolor[HTML]{FFCCC9} 
21  &  Ld Image  & 20:57:43:652  & 705594  & 12032  & 10464  & 0  & \%SysWOW64\%\textbackslash{}cscript.exe  &  & \%SysWOW64\%\textbackslash{}mswsock.dll  \\
\rowcolor[HTML]{9AFF99} 
22  &  Tr Create  & \multicolumn{1}{l}{\cellcolor[HTML]{9AFF99}20:57:44:210} & \multicolumn{1}{l}{\cellcolor[HTML]{9AFF99}707985} & 12032  & \multicolumn{1}{l}{\cellcolor[HTML]{9AFF99}10464} & \multicolumn{1}{l}{\cellcolor[HTML]{9AFF99}2844} & \%SysWOW64\%\textbackslash{}cscript.exe  &  &  \\
\rowcolor[HTML]{FFCCC9} 
23  &  Ld Image  & 20:57:44:226  & 708195  & 12032  & 10464  & 0  & \%SysWOW64\%\textbackslash{}cscript.exe  &  & \%SysWOW64\%\textbackslash{}urlmon.dll  \\
\rowcolor[HTML]{FFCCC9} 
24  &  Ld Image  & 20:57:44:232  & 708207  & 12032  & 10464  & 0  & \%SysWOW64\%\textbackslash{}cscript.exe  &  & \%SysWOW64\%\textbackslash{}virtdisk.dll  \\
\rowcolor[HTML]{96FFFB} 
25  &  Tr Exit  & \multicolumn{1}{l}{\cellcolor[HTML]{96FFFB}20:57:44:249} & \multicolumn{1}{l}{\cellcolor[HTML]{96FFFB}708455} & 12032  & \multicolumn{1}{l}{\cellcolor[HTML]{96FFFB}10464} & \multicolumn{1}{l}{\cellcolor[HTML]{96FFFB}2844} & \%SysWOW64\%\textbackslash{}cscript.exe  &  &  
\end{tabular}
}
\caption{\thesystem artifacts on process activities.}
\label{tab:malawre_activity_trace_process}
\end{subtable}\par


\begin{subtable}{0.9\textwidth}
\centering
\resizebox{\textwidth}{!}{ 
\begin{tabular}{cccccccll}
\hline
\multicolumn{1}{|c|}{\#} & \multicolumn{1}{|c|}{Operation} & \multicolumn{1}{c|}{Time} & \multicolumn{1}{c|}{Duration} & \multicolumn{1}{c|}{Global \#} & \multicolumn{1}{c|}{PID} & \multicolumn{1}{c|}{TID} & \multicolumn{1}{c|}{Image path}  & \multicolumn{1}{c|}{File Path}   \\ \hline
\rowcolor[HTML]{96FFFB} 
1  &  IRP\_Create  & 20:51:45:636  & 69  & 183856  & 10092  & 1504  & \%MSOffice\%\textbackslash{}EXCEL.EXE & C:\textbackslash{}Users\textbackslash{}grace\textbackslash{}Downloads\textbackslash{}ORDER SHEET \& SPEC.xlsm   \\
\rowcolor[HTML]{FFCCC9} 
2  &  IRP\_Write  & 20:51:48:590  & 1757  & 211615  & 10092  & 3620  & \%MSOffice\%\textbackslash{}EXCEL.EXE & C:\textbackslash{}Users\textbackslash{}grace\textbackslash{}AppData\textbackslash{}Local...\textbackslash{}Temp\textbackslash{}DED9E0FE.xlsm  \\
\rowcolor[HTML]{FFCCC9} 
3  &  IRP\_Write  & 20:51:48:755  & 448  & 213650  & 1612  & 12240  & \%MSOffice\%\textbackslash{}EXCEL.EXE & C:\textbackslash{}Users\textbackslash{}grace\textbackslash{}AppData\textbackslash{}Local\textbackslash{}Packages\textbackslash{}...\textbackslash{}AC\textbackslash{}Temp\textbackslash{}560D285B.emf \\
\rowcolor[HTML]{FFCCC9} 
4  &  IRP\_Write  & 20:52:13:024  & 501  & 296839  & 10092  & 1504  & \%MSOffice\%\textbackslash{}EXCEL.EXE & C:\textbackslash{}Users\textbackslash{}grace\textbackslash{}AppData\textbackslash{}Local\textbackslash{}Microsoft\textbackslash{}...\textbackslash{}1983A0E7.png  \\
\rowcolor[HTML]{FFCCC9} 
5  &  IRP\_Write  & 20:52:16:156  & 468  & 345357  & 10092  & 1504  & \%MSOffice\%\textbackslash{}EXCEL.EXE & C:\textbackslash{}Users\textbackslash{}grace\textbackslash{}AppData\textbackslash{}Local\textbackslash{}Microsoft\textbackslash{}Windows\textbackslash{}...\textbackslash{}1959A28D.emf \\
\rowcolor[HTML]{FFCCC9} 
6  &  IRP\_Write  & 20:52:16:230  & 886  & 345950  & 10092  & 1504  & \%MSOffice\%\textbackslash{}EXCEL.EXE & C:\textbackslash{}Users\textbackslash{}grace\textbackslash{}AppData\textbackslash{}Local\textbackslash{}Temp\textbackslash{}q  \\
\rowcolor[HTML]{FFCCC9} 
7  &  IRP\_Write  & 20:52:16:270  & 551  & 346113  & 10092  & 1504  & \%MSOffice\%\textbackslash{}EXCEL.EXE & C:\textbackslash{}Users\textbackslash{}grace\textbackslash{}AppData\textbackslash{}Local\textbackslash{}Temp\textbackslash{}xx  \\
\rowcolor[HTML]{FFCE93} 
8  &  IRP\_Set\_Information  & 20:52:16:932  & 748  & 355451  & 12148  & 8056  & \%SysWOW64\%\textbackslash{}wscript.exe & C:\textbackslash{}Users\textbackslash{}grace\textbackslash{}AppData\textbackslash{}Local\textbackslash{}Temp\textbackslash{}xx  \\
\rowcolor[HTML]{CBCEFB} 
9  &  IRP\_Close  & 20:52:16:933  & 17  & 355455  & 12148  & 8056  & \%SysWOW64\%\textbackslash{}wscript.exe & C:\textbackslash{}Users\textbackslash{}grace\textbackslash{}AppData\textbackslash{}Local\textbackslash{}Temp\textbackslash{}xx.vbs  \\
\rowcolor[HTML]{FFCCC9} 
10  &  IRP\_Write  & 20:53:05:692  & 385  & 407107  & 10092  & 1504  & \%MSOffice\%\textbackslash{}EXCEL.EXE & C:\textbackslash{}ProgramData\textbackslash{}asc.txt:script1.vbs    \\
\rowcolor[HTML]{9AFF99} 
11  &  IRP\_Read  & 20:57:42:541  & 73  & 679583  & 10464  & 2584  & \%SysWOW64\%\textbackslash{}cscript.exe & C:\textbackslash{}Users\textbackslash{}grace\textbackslash{}AppData\textbackslash{}Local\textbackslash{}Temp\textbackslash{}xx.vbs  \\
\rowcolor[HTML]{FFCCC9} 
12  &  IRP\_Write  & 20:57:44:237  & 3432  & 708409  & 10464  & 2844  & \%SysWOW64\%\textbackslash{}cscript.exe & C:\textbackslash{}ProgramData\textbackslash{}Podaliri4.exe   \\
\rowcolor[HTML]{CBCEFB} 
13  &  IRP\_Close  & 20:57:44:248  & 20  & 708591  & 10464  & 2844  & \%SysWOW64\%\textbackslash{}cscript.exe & C:\textbackslash{}ProgramData\textbackslash{}Podaliri4.exe   
\end{tabular}
}
\caption{\thesystem artifacts on filesystem activities.}
\label{tab:malawre_activity_trace_file}
\end{subtable}\par
\label{tab:malawre_activity_trace}
\end{table*}

\section{Case Studies}
\label{sec:measurement}
\noindent \thesystem is designed to be portable, low artifact, and robust against common evasion mechanisms. In this section, we provide two case studies
to illustrate how the engine can achieve these goals under different settings and deployment scenarios. 
In case study 1, we deployed \thesystem in a baremetal malware analysis environment, aiming to address common fingerprinting techniques in the wild. The analysis is based on 79,544 malware samples. In the second case study, 
we deploy \thesystem as an in-cloud deception-based threat intelligence service with the intent to collect real-world artifacts on how adversaries exploit vulnerable machines and how they use those hijacked. 




\subsection{Case Study 1: A Baremetal-assisted Analysis Infrastructure} 
\label{subsec:usecase1}


Virtualized environments have been used significantly in prior research~\cite{kharaz2016unveil,9230384,10.1145/3642974.3652280,10.1007/978-3-319-60876-1_5} to analyze malicious code due to the ease of deployment, scalability, isolation, and hardware utilization. However, this approach often comes with important visibility costs. In particular, environmentally sensitive samples launch several forms of fingerprinting checks to analyze the target host. They rarely load their actual malicious payloads in these environments, making analysis and reverse engineering a challenging task.

\begin{table}[!ht]
\caption{A subset of malware samples with the corresponding labels analyzed by \thesystem-enabled threat analysis platform. The dataset contains four major malware types, 941 malware families, and 3,596 malware variants.}
     \label{tab:payload_distribution}
\small
\center{\sffamily
\resizebox{\columnwidth}{!}{
\begin{tabular}{lrr}
      \toprule
        \textbf{Malware} & \textbf{Families-Occurrences} & \textbf{ Variants}\\
    \midrule
     \textbf{Ransomware}  &\textbf{64 - 472 (1.18\%)}\\
         \hspace{3mm}StopCrypt   &  253 (0.63\%) & 48  \\
         \hspace{3mm}Gandcrab   &  23 (0.06\%) & 13  \\
         \hspace{3mm}Lockbit   &  20 (0.05\%) & 10  \\
         \hspace{3mm}Hive   &  19 (0.05\%) & 6  \\
         \hspace{3mm}MedusaLocker   &  15 (0.04\%) & 1  \\
         \hspace{3mm}Filecoder   &  14 (0.03\%) & 8  \\
         \hspace{3mm}Others   &  58 - 128 (0.32\%) & 71 \\

    \textbf{RAT}&\textbf{ 602 - 31,158 (77.79\%)}\\
         \hspace{3mm}Sabsik   &  3,474 (8.67\%) & 8  \\
         \hspace{3mm}Berbew   &  3,116 (7.78\%) & 5  \\
         \hspace{3mm}AgentTesla   &  2,362 (5.90\%) & 457  \\
         \hspace{3mm}Wacatac   &  2,009 (5.01\%) & 3  \\
         \hspace{3mm}Vindor   &  1,175 (2.93\%) & 3  \\
         \hspace{3mm}RpcDcom based   &  1,069 (2.67\%) & 1  \\
         \hspace{3mm}Others   &  596 - 17,953 (44.83\%) & 2,495  \\

    \textbf{PUP} &\textbf{163 - 1,322 (3.30\%)}\\
         \hspace{3mm}Zbot   &  92 (0.23\%) & 8  \\
         \hspace{3mm}Infostealer   &  83 (0.21\%) & 1  \\
         \hspace{3mm}KuaiZip   &  80 (0.20\%) & 1  \\
         \hspace{3mm}AutoKMS   &  75 (0.19\%) & 3  \\
         \hspace{3mm}Ymacco   &  72 (0.18\%) & 16  \\
         \hspace{3mm}DarkStealer   &  67 (0.17\%) & 2  \\
         \hspace{3mm}Others   &  157 - 853 (2.12\%) & 186  \\

    \textbf{Self Replicating Malware} &\textbf{112 - 7,098 (17.72\%)}\\
         \hspace{3mm}Sfone   &  982 (2.45\%) & 2  \\
         \hspace{3mm}Ganelp   &  759 (1.89\%) & 7  \\
         \hspace{3mm}Viking   &  752 (1.87\%) & 16  \\
         \hspace{3mm}Xolxo   &  707 (1.76\%) & 1  \\
         \hspace{3mm}Autorun   &  595 (1.48\%) & 13  \\
         \hspace{3mm}Vobfus   &  492 (1.23\%) & 50  \\
         \hspace{3mm}Others   &  106 - 2,811 (7.02\%) & 161  \\
    
    \midrule
        \textbf{Total}    & \textbf{941 - 40,050}      & \textbf{3,596}   \\
       \bottomrule
 \end{tabular}}}
\end{table}

\noindent In this case study, we study the feasibility of \thesystem as a low-artifact engine and build a scalable baremetal analysis environment to collect artifacts about modern malware payloads. 
The core motivation behind this case study was to answer the question of how to build threat monitoring platforms that can satisfy common fingerprinting techniques often used in modern malware samples. We started by asking how much visibility could be gained by running the evasive samples in the baremetal analysis environment.

\noindent This experiment was performed by executing 79,544 malware samples from across 941 malware families and analyzing the results. Table \ref{tab:payload_distribution} shows the distribution of malware samples across different families for the 40,050 labels obtained. Each sample was executed on both the virtual and baremetal machines for five (5) minutes~\cite{kuechlerndss21}, during which run-time behavior artifacts were collected. 
The infrastructure used is equipped with high-bandwidth network switches to enable transferring Windows 10 images to the device in each run. That is, in each run, a new \thesystem-enabled OS image was distributed and loaded into each machine. 
The baremetal machines had similar processing, memory, and storage specifications.

\begin{figure*}[!ht]
    \begin{subfigure}[b]{0.66\columnwidth}
         \includegraphics[width=1.0\linewidth]{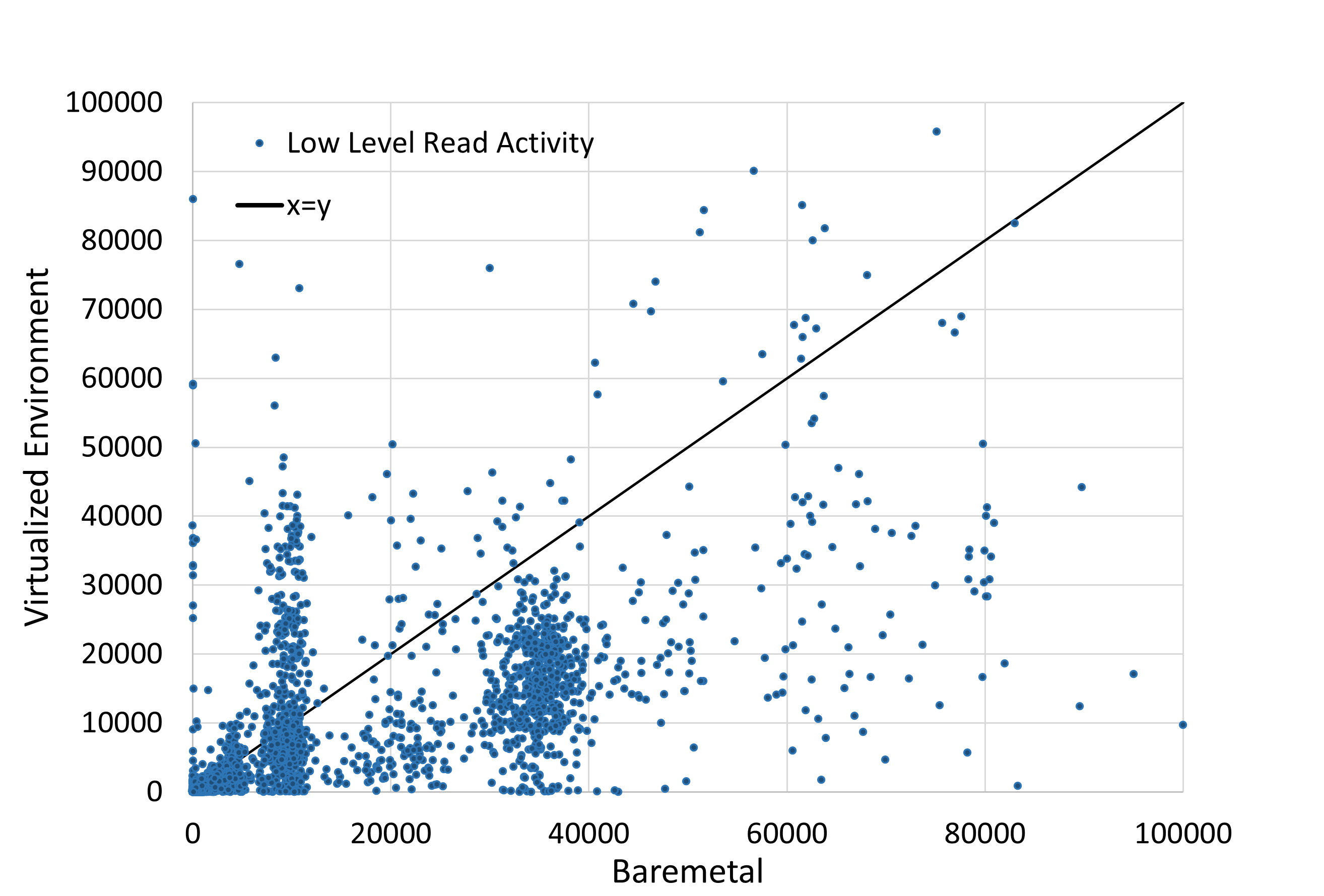}
         \captionsetup{font=scriptsize}
         \caption{Read Activity}
         \label{fig:IRP_MJ_READ}
     \end{subfigure}
     \hfill
     \begin{subfigure}[b]{0.66\columnwidth}
         \includegraphics[width=1.0\linewidth]{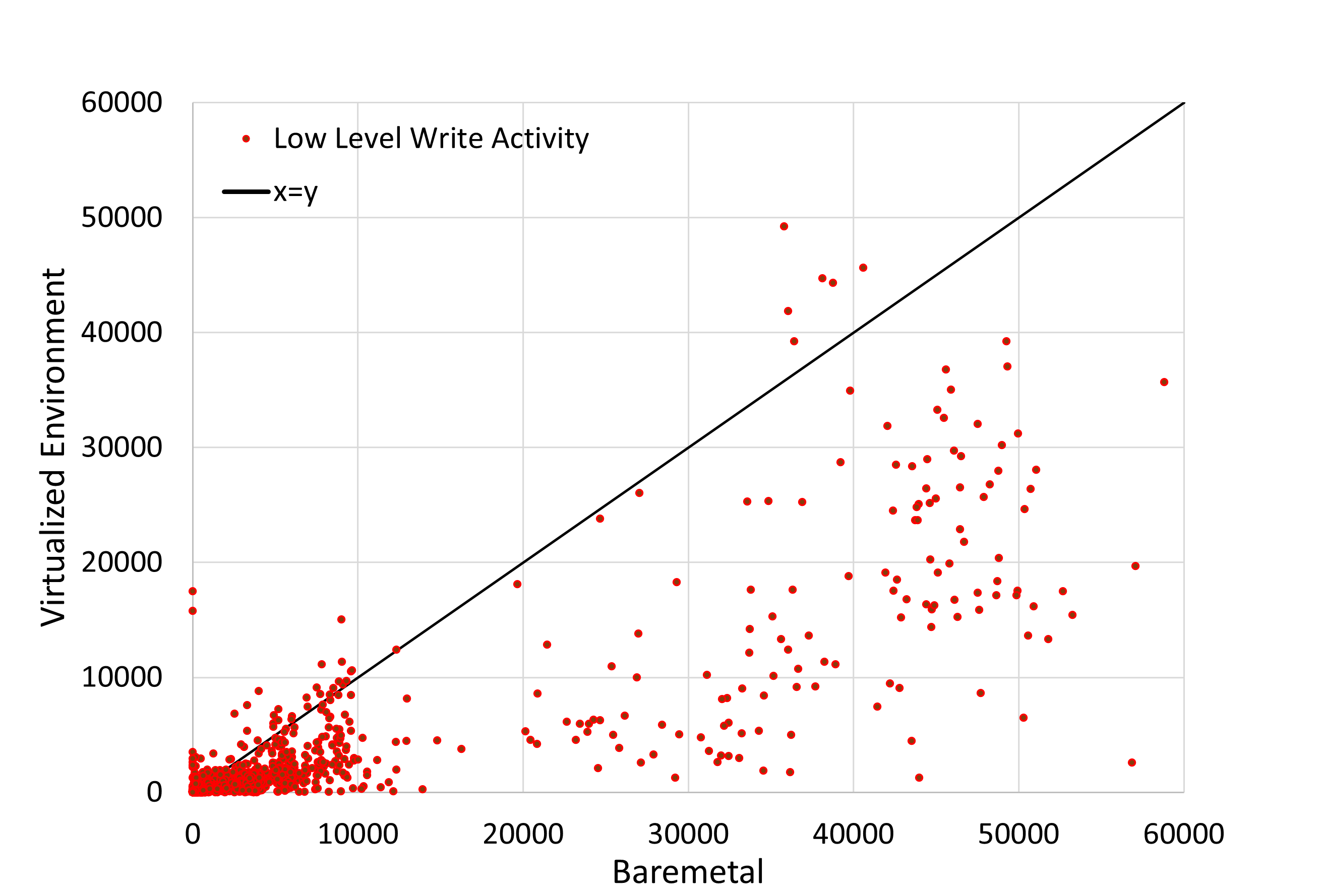}
         \captionsetup{font=scriptsize}
         \caption{Write Activity}
         \label{fig:IRP_MJ_WRITE}
     \end{subfigure}
     \hfill
        \begin{subfigure}[b]{0.66\columnwidth}
         \includegraphics[width=1.0\linewidth]{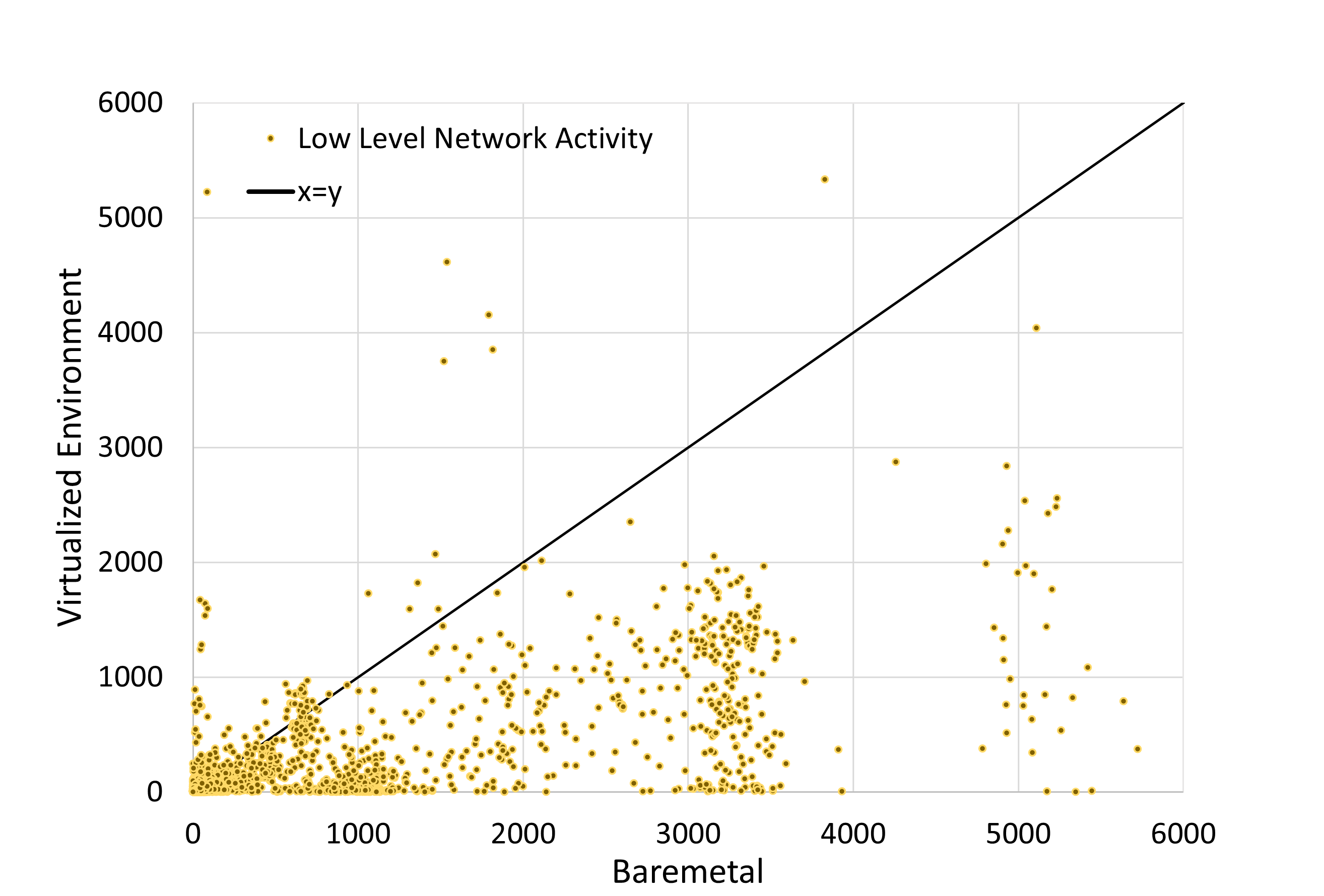}
         \captionsetup{font=scriptsize}
         \caption{Cached-File Open Activity}
         \label{fig:IRP_MJ_NETWORK_QUERY_OPEN}
     \end{subfigure}
     \hfill
     \begin{subfigure}[b]{0.66\columnwidth}
         \includegraphics[width=1.0\linewidth]{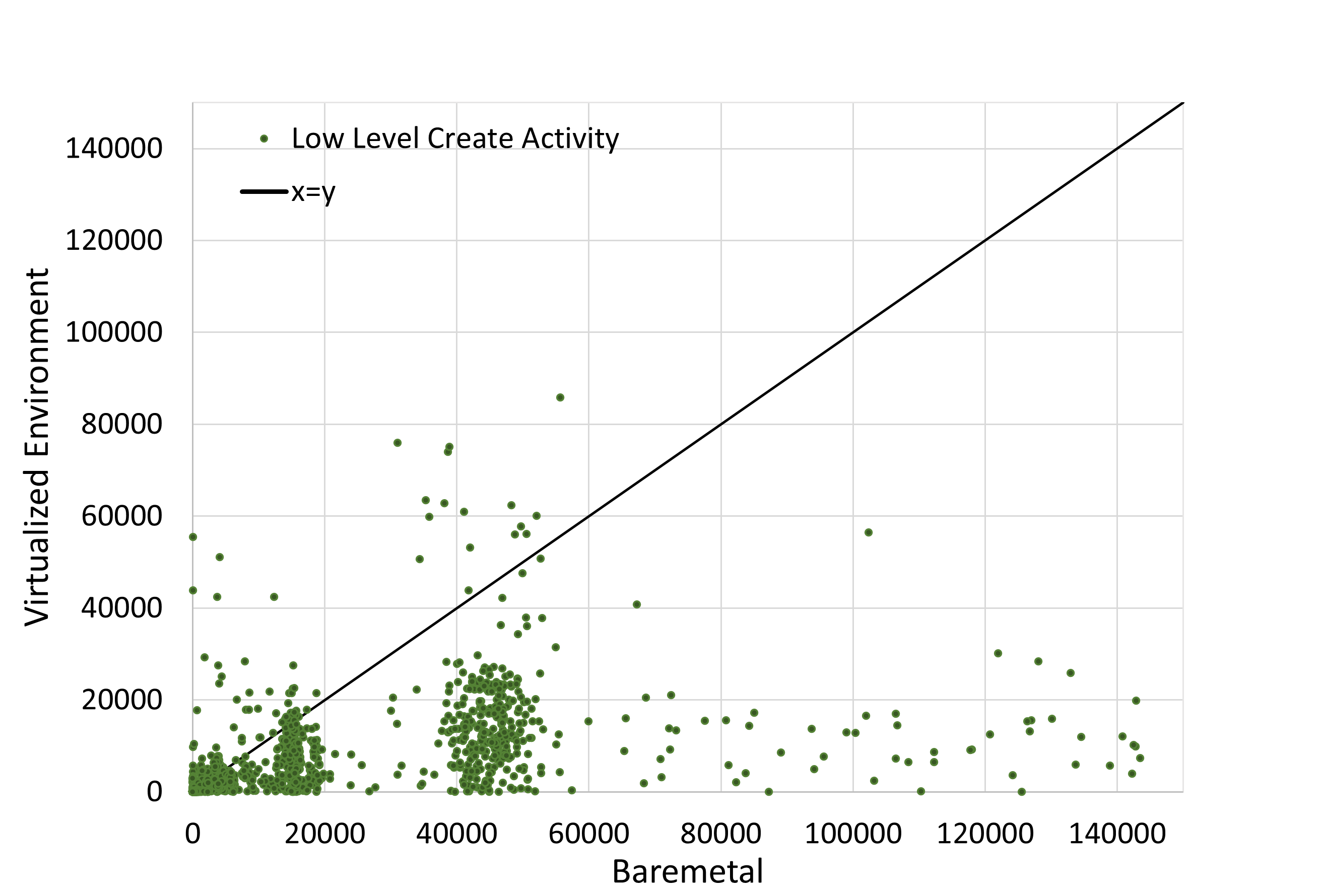}
         \captionsetup{font=scriptsize}
         \caption{File Create Activity}
         \label{fig:IRP_MJ_CREATE}
     \end{subfigure}
     \hfill
     \begin{subfigure}[b]{0.66\columnwidth}
         \includegraphics[width=1.0\linewidth]{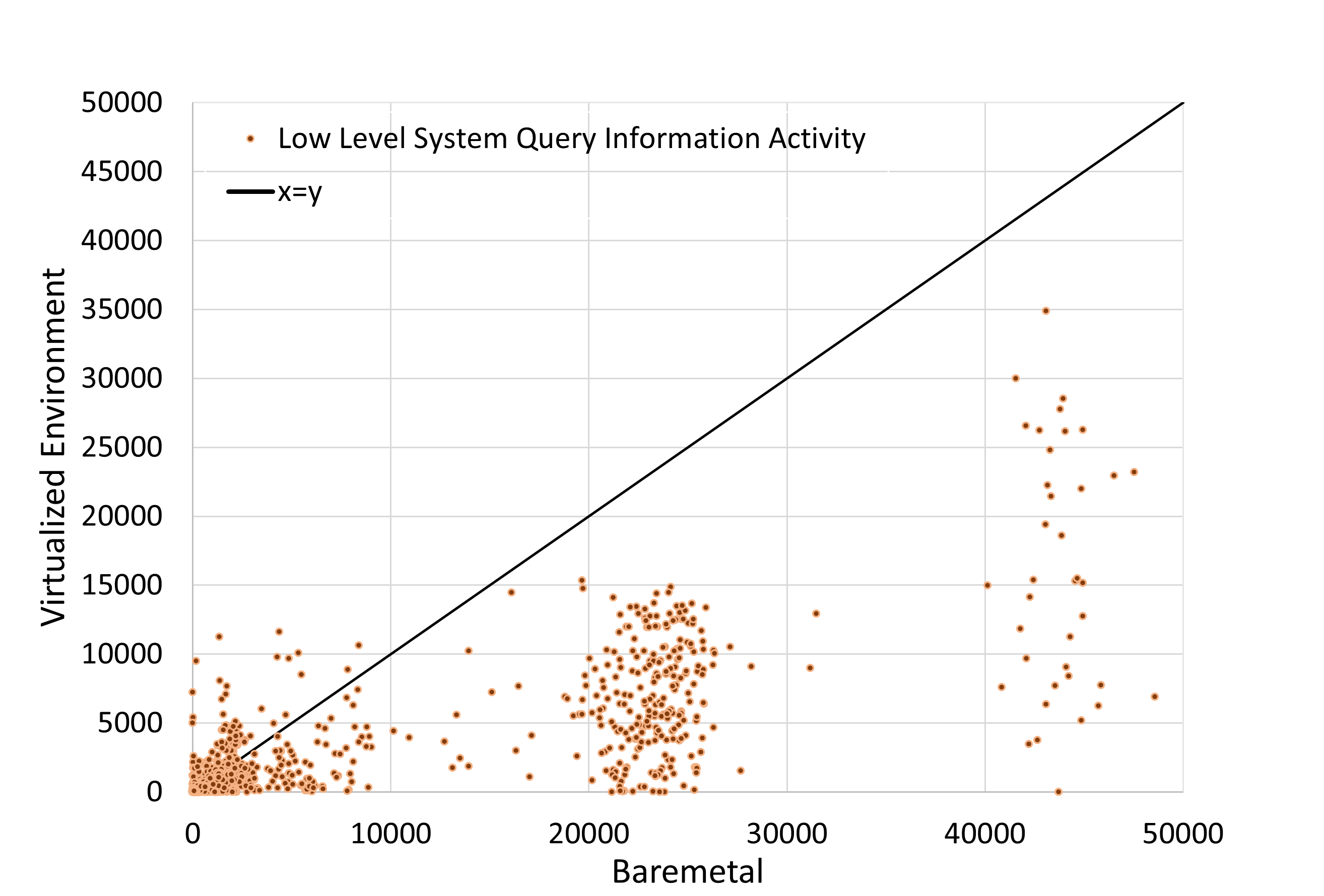}
         \captionsetup{font=scriptsize}
         \caption{System Query Information Activity}
         \label{fig:IRP_MJ_QUERY_INFORMATION}
     \end{subfigure}
   \hfill
     \begin{subfigure}[b]{0.66\columnwidth}
         \includegraphics[width=1.0\linewidth]{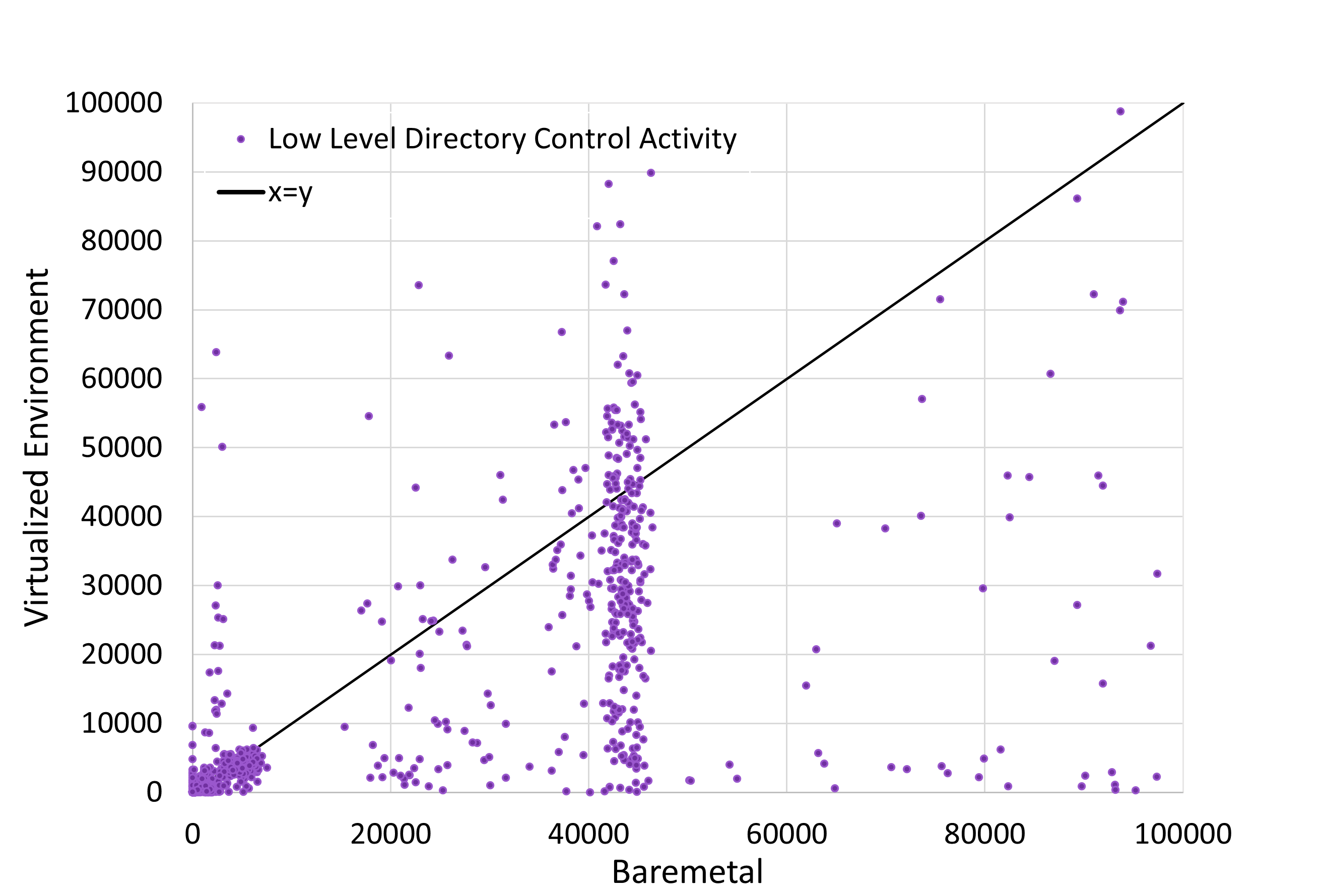}
         \captionsetup{font=scriptsize}
         \caption{Directory Control Activity}
         \label{fig:IRP_MJ_DIRECTORY_CONTROL}
     \end{subfigure}
        \caption{File system activities in \thesystem-enabled machines in baremetal and virtual environments. The results show that malware samples manifest more activities in the baremetal environment, resulting in more dropped files and execution payloads.}
        \label{fig:baremetal_vs_vm}
\end{figure*}

\noindent Figure~\ref{fig:baremetal_vs_vm} shows the summary of our analysis. While \thesystem records 45 various low-level OS operations, we have provided six major operations to show the run-time behavior of a given sample. As shown, while there are samples with a similar number of key operations in both environments, there are still a large number of samples that show significantly more activities in baremetal compared to virtualized environments. For instance, the number of write operations (Figure~\ref{fig:IRP_MJ_WRITE}) is almost 64\% more than the ones in a virtualized environment for 40,050 samples. 
The difference in the type of dropped payloads in the two environments suggests that the malware might have taken different execution paths in the two environments. 
We noticed the download of exe files, as a result of launching the malware, had a significant decrease (93.9\%) in the virtualized environment.
JavaScript files (JS) are the second most common type of malware payload downloaded, accounting for over 12.68\% of all payloads. This is most likely due to the growing popularity of web-based attacks, as JS is the most often used scripting language on websites~\cite{github2023octoverse}. PDF documents are the most common type of non-executable payload downloaded, accounting for over 30\% of all non-executable document payloads. The prevalence of PDFs as a document-sharing format is a probable cause, as they can serve as a vehicle for embedding and executing malicious code upon opening~\cite{maiorca2019towards, issakhani2022pdf}. Table \ref{tab:dropped_extensions} in the appendix shows a more comprehensive version of the dropped files in the experiments. 

\noindent We performed an analysis of the fingerprinting checks used by samples in the dataset that cause the differences in the execution traces of \thesystem-enabled machines in baremetal and virtual environments. Our analysis shows that 1,187 samples from at least five malware families call Windows Management instrumentation API to get the actual hardware-level driver information before loading their actual payloads. Direct access to CPU clocks, bios, and timing checks have also been seen across various malware families in the test, as shown in table \ref{tbl:fingerprinting}.




\begin{table}[!h]
    \centering
    \caption{A subset of environment fingerprinting checks satisfied by \thesystem, compared to other core technologies.\faCheck=Satisfied, \faTimes=Failed
    }
    \footnotesize
     \resizebox{0.99\columnwidth}{!}{
    \begin{tabular}{lrcccl}
    \toprule
        \textbf{FP Check}  & \textbf{Samples}  & \textbf{Baremetal-Based} & \textbf{VM-Based}& \textbf{Top Families} \\ \midrule
        calls-wmi & 1,187 &\faCheck& \faTimes& AgentTesla, Sabsik, Redline, Leone, Znyon \\
        direct-cpu-clock-access & 9,254 &\faCheck & \faTimes& AgentTesla, Sabsik, Wacatac, Woreflint, FormBook \\
        checks-bios & 6,859& \faCheck& \faTimes& AgentTesla, Leone, Znyon, Formbook, Wacatac \\
        GetTickCount & 1,288  &\faCheck  & \faTimes& AgentTesla, Sabsik, Woreflint, Wacatac, FormBook \\
        
         \bottomrule
      
    \end{tabular}
    }
    \label{tbl:fingerprinting}
    \vspace{-0.3cm}
\end{table}

\noindent An immediate conclusion from this study is that a low-artifact analysis environment that can satisfy the hardware-based fingerprinting is critical to achieving more visibility in today’s attack landscape. 
That said, a baremetal-assisted solution should not be considered as an alternative solution to the current hypervisor-based solution that can offer significant reverse engineering flexibility. Instead, \thesystem-enabled malware analysis can serve as the first line of defense in malicious code analysis by satisfying fingerprinting checks that are impossible or very costly in other defense systems. Furthermore, the results also show the need for a more effective analysis environment for the research community that forces a trade-off between achieving more fine-grained visibility while leaving minimally detectable artifacts on the analysis machines.

\subsection{Case Study 2: \thesystem for Distributed Threat Intelligence}
\label{subsec:usecase3}

\noindent 
Adversaries are continuously becoming more effective in launching low-profile attacks on critical systems. 
The question that arises is how to collect relevant evidence about these attacks and their strategies at the early stages
of their operations. Lack of proper evidence on attack strategies can potentially put defenders in a highly disadvantaged
position because understanding attack tactics and lateral movements becomes complex. While tools and services such as Event Tracing Windows (ETW) offer critical insights on systems, performing root cause analysis and formulating a proper response often requires more fine-grained data. In this case study, we developed a distributed deception-based threat intelligence infrastructure, using \thesystem to gather evidence on real-world attacks on an intentionally exposed service.


\noindent 
To run this experiment, we deployed \thesystem-enabled cloud instances on Windows 11 x64 with 2vCPUs. Run-time traces were collected via a high-privilege triage engine to manage collection, packing, and subsequent posting to our remote servers. By design, we made our vulnerable systems detectable by automated scanners that actively scan network-based services for weak Remote Desktop Protocol (RDP) credentials. Intrusions were detected by monitoring a subset of high-valued files; once touched, we received an immediate notification of a successful intrusion. In an effort to lessen the likelihood of encountering the same attack campaigns repeatedly, we reset the servers and assign a random IP address after each successful compromise.

\begin{center}
    \begin{figure}[!ht]
    \includegraphics[width=\columnwidth]{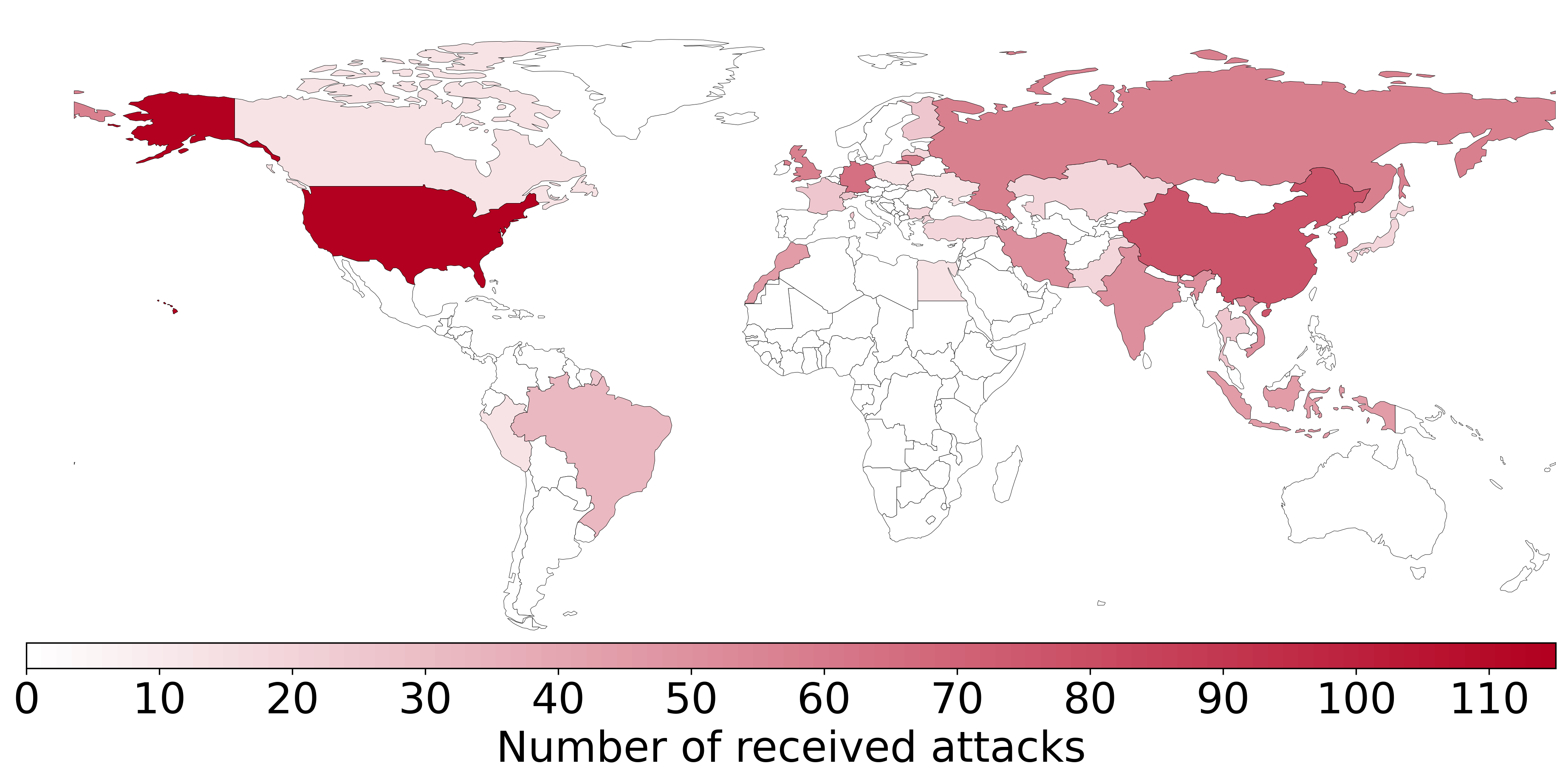}
    \caption{Distribution of the origin of successful attacks. While US-based attacks are the largest by volume, we observed that most of them are being abused as proxy nodes.}
    \label{fig:use_case3_intrusions_by_country}
    \end{figure}
\end{center}

\noindent We executed the experiments for a duration of 46 days, from December 19, 2023, to February 3, 2024. During this time, there were interactions with the environment from 398 unique IP addresses originating within 175 geographical locations. 
Figure~\ref{fig:use_case3_intrusions_by_country} shows the distribution of successful attack origins during the experiment timeline.
Our data shows that the exposed, vulnerable services were discovered on average 6 hours after publishing the instances, and successful exploitation occurred on average 13 hours after publishing the decoy server.
We noticed that the high-valued files were opened in 32 of the experiments. 
Privilege escalation was inferred in three cases where we observed that files were written in the system32 folder (an operation requiring elevated permissions) using the logs generated by the forensic engine. In 51 experiments, at least one file was retrieved from remote servers and dropped on the machines. In particular, \thesystem collected 196,774 dropped files, including 1,221 executable files, 4,586 DLL files, 401 shell scripts, 14,706 program source code files, 235 installers, and 438 digital certificates.

%


\noindent We identified attack strategies that allowed adversaries to monetize the compromised machine or run code, which facilitated further exploitation or persistence.  In the following, we describe some of the attack strategies we observed.

\noindent \textbf{Using the compromised machine as a proxy.} 
Traffic jacking is the process of converting a compromised machine into a proxy server and renting the server's bandwidth for adversarial purposes~\cite{mehanna2024free}. To initiate traffic jacking, we observed that the adversary dropped an executable that loaded 183 DLLs into the memory, including several cryptographic and socket communication libraries. 
Once loaded, the process established a connection with a remote server to schedule tasks. Forensics trace data from \thesystem shows that the parent process initiated hundreds of threads on the infected machine. The machine established a large number of TCP connections with hundreds of remote IP addresses to relay traffic. 
The adversary ran the malicious process in user-mode, and the forensic data captured by \thesystem did not indicate any traces of establishing virtual interfaces for relaying packets. Our analysis suggests that the compromised machine was indeed converted to an exit node in this experiment.

\noindent \textbf{Account Creation and Modification.} 
In 15 of the exploitations, we observed indications of account creation and privilege escalation attempts using \texttt{net user} and \texttt{net localgroup} to gain persistent access. 
Another popular operation was group enumeration (e.g., WMIC Group where "SID = 'X' Get Name /Value | Find "=") to identify group memberships, specifically for admins and remote desktop users to potentially manipulate user privileges. 

\noindent \textbf{Backup Erasure}
We observed a pattern across 12 incidents where Powershell scripts attempted to delete the Volume Shadow Copies and backup catalog. We observed that adversaries were using commands such as 'vssadmin delete shadows /all /quiet', 'wbadmin delete catalog -quiet', or 'wmic shadowcopy delete' to prevent the recovery of data, making it more difficult to restore from backup after an attack. This practice is common among adversaries to either force the victim to pay a ransom fee or destroy the system logging evidence to complicate incident analysis.

\noindent \textbf{Persistence Threats.} We observed several attempts among adversaries to stay persistent on the exploited machines. 
We also observed 5 cases where the adversaries tried to modify the password policies to set passwords to never expire (e.g., net accounts /maxpwge:unlimited), making it easier for adversaries to maintain access. 
In 10 incidents, we observed listing and manipulation queries where adversaries attempted to add or modify registry entries to hide user accounts from the login screen, making the accounts less noticeable to the casual user.
In 5 cases, we observed no significant operation during the initial compromise. However, we observed setting scheduled tasks for future logins. In particular, we observed the use of \texttt{schtasks} /\texttt{create} to create tasks that execute at specific times or system events.

\section{Discussion}
\label{Discussion}

\noindent In this paper, we aimed to show that modern threat analysis highly depends on defining robust forensics engines that force a reliable trade-off between providing fine-gained behavioral insights and minimizing detectable artifacts. We posit that a solution that can achieve these two goals has several use cases on the defense side. As a first step in this direction, we defined two main application scenarios for \thesystem and showed how it can add a new lens to the analysis and detection of security incidents. We summarize our main findings along the following four points:

\noindent \textbf{More Visibility over Malware Behavior:} While prior work shows modern malware samples are getting less sensitive to virtualized environment~\cite{kuechlerndss21}, our data suggests that such environmental checks are still prevalent across different RAT, droppers, and PUP samples.
We observed that 64\% (22,021) of the samples with file write activities had at least 70\% more dropped or modified files in \thesystem compared to cases in virtualized environments. We acknowledge that code coverage can have various definitions in the context of program analysis. However, the fact that more filesystem activities were observed in head-to-head experiments and a larger number of payloads were delivered on average suggests that the malware sample was designed to take a different path in environments close to real-world settings. In a broader context, this procedure can be augmented with static analysis approaches to systemically analyze malicious code more deeply. 
We believe systems similar to \thesystem should be more accessible to threat researchers and malware analysts for further investigations in malicious code analysis due to the scarcity of real-world run-time behavior datasets on baremetal systems.




\noindent \textbf{\thesystem in a Broader Context.}
In this work, we aimed to answer how to improve behavioral visibility in modern evasive malicious code attacks while making them more robust to evasion. That is a critical question in malicious code analysis and the generalizability of the results to other analysis engines and sandboxes is important. For instance, in case study 1, we showed that our baremetal solution works more effectively in comparison to the virtualized environment in satisfying almost all debugging and VM-based checks. We expect similar results if the experiments are conducted in other virtualization technologies (e.g., hypervisor-based). 
In particular, we observed that it is quite common across modern attacks to call Windows Management Instrumentation API in order to gain information on system hardware, applications, networks, devices, and other managed components.
While manipulating the return values in WMI API to report fake responses is possible, it often leads to serious system crashes if the system has to operate for a longer period (case study 2) because the healthy operation of many system drivers depends on the responses received from the WMI API, and a fake response can easily cause a system crash. Unfortunately, 
hypervisor-based methods are still vulnerable to this fingerprinting method, although they offer significant freedom for other forms of analysis. 

\noindent \textbf{Unique Artifact Catalogs.}
The dataset we created in this experiment is the largest catalog of threat artifacts collected from a baremetal environment. While there are some available artifacts~\cite{ShieldFS, Malrec, kharaz2016unveil} that helped other work to have basic benchmarks, we are not aware of any large-scale run-time baremetal artifacts from prior work or commercial sandboxes on different forms of malware samples. 

\noindent The high cost of building robust solutions and the scarcity of high-quality data can potentially impact the development of new data-driven defenses in this dynamic ecosystem where collecting evidence on emerging trends is critical. Lack of access to malicious data and the inherent complexity of defining a robust defense mechanism make it extremely difficult to establish a common ground for comparison. 
Given the ability to create a usable, timely, and longitudinal database of threat artifacts, we propose a set of research applications that would enable lower-cost threat intelligence. For instance, having longitudinal access to the artifacts of real-world attacks can be useful in testing new defense tools, studying attack evolution, and evaluating defense models with reference datasets.

\noindent \textbf{Towards More Robust Cyber Deception Technologies.}
The ability to deploy \thesystem in different computing environments, including baremetal systems, end-points, and cloud environments, allows the deployment of a low-artifact deception-based threat intelligence infrastructure on the Internet. That is, deploying \thesystem-enabled machines will serve as distributed threat intelligence sensors, offering a unique opportunity to collect evidence on new attacks. Consider the recent attack on the MOVEit transfer service as an example~\cite{cisco}. 
A large-scale honeypot enabled by \thesystem could offer significant visibility on who first started the attack, how the attack took place, and a chronological order of the steps taken to perform the attack. As a next step, we plan to systematize the development of a vulnerability insertion module when a zero-day vulnerability is announced to measure the scope and scale, the evolution of the payloads and attacks, as well as the adversarial campaigns behind the attacks.

\subsection{Limitations}

\noindent Section~\ref{sec:measurement} demonstrates that \thesystem achieves practical and useful detection results on a large, real-world dataset. Unfortunately, adversaries continuously observe defensive advances and adapt their attacks accordingly. In the following, we discuss the limitations of \thesystem and potential evasion strategies.


\noindent \textbf{Delay Injection.} In case study 1, injecting a significant delay before loading the malicious payload is another method to bypass detection. The experiments were designed to record the behavior for five minutes, so if the malicious code stays dormant for a long period of time, \thesystem cannot record actions associated with the malicious activity. Note that these limitations are not specific to \thesystem, and almost any dynamic analysis system may be impacted in some way by these evasion methods. Prior work has discussed this evasion mechanism. Note that incorporating these techniques can also potentially make detection easier in static analysis since these approaches require calling specific functions in the operating system. The presence of these mechanisms in the initial binary is being used currently by malware defense solutions to identify suspicious binaries in the wild. 

\noindent \textbf{Trusted Computing Model.}
\thesystem operates at the kernel level. If the target malicious code is a kernel-level attack, it can potentially thwart some of the hooks \thesystem uses to monitor run-time behavior. Analyzing high-privilege attacks is out of the scope of this paper. These attacks can be better analyzed in a hypervisor-based environment where debugging kernel-level code is possible. We assume that the trusted computing base includes the display module, OS kernel, and underlying software and hardware stack. Therefore, we can safely assume that the components of the system are free of malicious code and that normal user-based access control prevents attackers from running malicious code with superuser privileges. This is a fair assumption considering the fact that most malicious operations (e.g., malware attacks and software vulnerability checking) are often initiated in the user mode.

\section{Related Work}
\label{RelatedWork}

\noindent Security research has explored various ways to understand the behavior of evasive code, contextualize the behavior, and predict possible adversarial action. In this section, we discuss prior work, mainly focusing on building tools and systems to collect and analyze threat artifacts. 


\noindent \textbf{Foreniscs Services:}
Many of the prior works utilized the built-in Windows OS, Event Tracing for Windows (ETW), for their security analysis.
For instance, 
Lee et al.~\cite{Lee2015ACSAC} used the cursory data provided by ETW to detect APTs, thus lowering the overhead for log storage. Other approaches by Hassan et al.~\cite{Hassaan2021IEEE} look at the problem using static analysis to identify structures from built-in operating system audits. Rapsheet's~\cite{hassan2020tactical} proposed approach involves the detection of APTs by condensing the acquired low-level data into graphical representations and subsequently comparing the observed steps with the publicly-accessible MITRE ATT\&CK knowledge base~\cite{mitre2023}. This knowledge base is curated by domain experts who analyze real-world APT attacks.
In fact, ETW is designed to primarily serve as a telemetry collection tool that gathers high-level event data. Although this is helpful, its use in forensic analysis is constrained because of the level and granularity of reports. Additionally, ETW is susceptible to fingerprinting and can be disabled by malware~\cite{Odzhan2020, Chester2020}. Contributions from Bakshi et al.~\cite{Bakshiosquery2023} built upon ETW to augment its ability to collect event activity logs. However, its ability to collect data at the granularity needed for a detailed low-level analysis still falls short~\cite{NUNES2019102365}. Lower-level artifacts such as monitoring file systems, networks, and processes are more useful in this context~\cite{NUNES2019102365}. 

\noindent At the kernel level, data can be collected using global operating system-level abstractions such as processes, files, and network interactions~\cite{Hollasch2022}. There have been several works in which the implementation of lower-level artifact collection is centered around the file system alone~\cite{kharaz2016unveil,scaife2016cryptolock,kharraz2017redemption,continella2016shieldfs,sgandurraelderan2016, ABBASIASC2022, Leecs2022}. This is advantageous for detecting changes to the filesystem, especially in cases of ransomware in which many files are modified during a relatively short window~\cite{kharaz2016unveil,sunieeechameleon2022,scaife2016cryptolock}. A further extension incorporating data collection at a process level is included in RwGuard~\cite{MehnazRwGuard2018} whenever it was determined that the interactions were suspicious. It approaches the problem by deploying a logger in the kernel space and adding a process monitoring aspect within the user space. Although the methodology is adept at collecting process information, it is still tightly coupled with their logger. Still in the kernel at a memory level, Shah et al.~\cite{shah2022memory} and Kara~\cite{KARAesa2023} proposed systems for detecting malware that resides in computer memory. After executing malware, they extract a memory dump that is further converted into an image for processing using computer vision and machine learning. 

\noindent One gap in this domain is that the core technology used in prior work to collect threat intelligence data either offered relevant but high-level insights about incidents or was primarily focused on one class of attacks (e.g., ransomware). 
To our knowledge, \thesystem is the only open-source solution that collects kernel-level data points (processes and threads, filesystem, registry) and generates GBs of data on the collected artifacts.

\noindent \textbf{Sandbox and behavioral Analysis.}
A drawback for any system collecting data in this adversarial space is its susceptibility to being fingerprinted~\cite{galloro2022systematical, deliaieee2020, or2019dynamic}. Advances in malware construction have seen the advent of anti-debugging and anti-sandboxing techniques to evade analysis and detection by security researchers trying to analyze malware behavior~\cite{deliadt2022,galloro2022systematical}. These techniques are principally designed to make analysis time-consuming and tedious for researchers. It has been observed that malware behaves differently when it perceives that the environment is sandboxed for analysis or executed in a virtualized environment~\cite{lee2021bypassing, kirat2011barebox, sunieee2011,kuechlerndss21, erin2021usenixsec, liu2022enhancing, LIU2022102613}. In some cases, the malware behaves benignly or does not execute if it detects a sandboxed environment~\cite{guibernau2020catch,miramirkhani2017spotless}. Proposals for anti-sandboxing techniques Liu et al.~\cite{LIU2022102613} build upon the works of Miramirkhani et al.~\cite{miramirkhani2017spotless} and Hu et al.~\cite{spensky2016phi} in creating believable system artifacts to mitigate system fingerprinting. 

\noindent As a technique for sidestepping the issue of virtualized systems being detected, several works have proposed the use of physical machines for the collection of forensic data~\cite{kirat2011barebox, spensky2016phi,kirat2014barecloud}. BareBox~\cite{kirat2011barebox} uses a restoration method in which snapshots of memory and disks are used simultaneously with parallel operating systems. BareCloud~\cite{kirat2014barecloud} uses a distributed model of disks, restoring them after each execution. LO-PHI~\cite{spensky2016phi}, like BareCloud uses machines that do not contain instrumentation for fingerprinting. 
\thesystem is designed to complement all these methods by being portable, allowing deployment of the \thesystem as an OS service in any computing environment.

\noindent \textbf{Fingerprinting and Evasive Techniques.}
Resistance to profiling in malware is presented in different forms where malicious authors develop a multitude of techniques to hide the true behavior of their applications whenever an analysis environment is detected. To this end, several studies~\cite{maffia2021longitudinal, galloro2022systematical, geng2024survey, nunes2022bane, yin2023method} have focused on understanding the techniques used to evade analysis. Maffia et al.~\cite{maffia2021longitudinal} investigate each technique, utilizing publicly available data to create an evasive program profiler to detect and circumvent evasive measures and collect statistical data on evasion methods. Similar research was carried out by Garollo et al.~\cite{galloro2022systematical} to produce statistically significant results on the link between malware families, the commonality of evasion techniques, and modifications to evasion techniques within families. They investigated evasive techniques in legitimate software and found that they are employed less frequently. Classification of malware by their evasive behaviors is studied by Yin and Nunes~\cite{yin2023method, nunes2022bane}. Nunes et al.~\cite{nunes2022bane} note that applications are classified as malicious simply based on evasive methods taken.

\section{Conclusions}
\label{Conclusions}

\noindent In this paper, we propose \thesystem, a low-artifact in-kernel forensics analysis system that aims to achieve improved visibility over the dynamics of the threat landscape. \thesystem is designed to optimize a balance between offering fine-grained visibility while minimizing detectable artifacts. 
We show these properties are critical to collecting all the relevant forensics events and conducting evidence-based threat characterization.
We evaluated \thesystem by running two case studies. In the first case study, we integrated \thesystem in a baremetal analysis environment and analyzed more than 79K malware samples. We collected GBs of artifacts on modern malware threats. 
In the second case study, we deployed the system as a deception-based threat intelligence, aiming to collect real-world threat artifacts. Our analysis showed that \thesystem was successful in collecting artifacts about hundreds of binaries and shell scripts developed and executed by adversaries on the threat infrastructure.


\bibliographystyle{plain}
\bibliography{bibliography}


\begin{table}[!ht]
 \caption{Distribution of a subset of 2.6 TBs of dataset submitted as artifacts in this submission. This represents 2\% of the artifacts. Each log contains traces by \thesystem from baremetal and the virtual environment.}
 \label{tab:malware_distribution_in_artifacts}
\small
\center{\sffamily
\begin{tabular}{lrr}
      \toprule
        \textbf{Malware} & \textbf{Occurrences} \\
    \midrule
     \textbf{Ransomware}  &\textbf{1 (0.13\%)}\\
        \hspace{3mm}StopCrypt & 1  (0.13\%) \\

    \textbf{RAT}&\textbf{669 (83.83\%)}\\
        \hspace{3mm}Banload & 1  (0.13\%) \\
        \hspace{3mm}CoinMiner & 35  (4.39\%) \\
        \hspace{3mm}CryptInject & 401  (50.25\%) \\
        \hspace{3mm}Emotet & 3  (0.38\%) \\
        \hspace{3mm}Farfli & 1  (0.13\%) \\
        \hspace{3mm}Glupteba & 1  (0.13\%) \\
        \hspace{3mm}Killav & 2  (0.25\%) \\
        \hspace{3mm}Musecador & 4  (0.50\%) \\
        \hspace{3mm}Phonzy & 1  (0.13\%) \\
        \hspace{3mm}Plyromt & 1  (0.13\%) \\
        \hspace{3mm}Pykspa & 1  (0.13\%) \\
        \hspace{3mm}Sabsik & 69  (8.65\%) \\
        \hspace{3mm}Trickbot & 1  (0.13\%) \\
        \hspace{3mm}Wabot & 18  (2.26\%) \\
        \hspace{3mm}Wacatac & 2  (0.25\%) \\
        \hspace{3mm}Ymacco & 128  (16.04\%) \\

    \textbf{PUP} &\textbf{2 (0.25\%)}\\
        \hspace{3mm}GameBox & 2 (0.25\%) \\

    \textbf{Self Replicating Malware} &\textbf{126 (15.79\%)}\\
        \hspace{3mm}Andriod & 5  (0.63\%) \\
        \hspace{3mm}Cambot & 6  (0.75\%) \\
        \hspace{3mm}Canbis & 5  (0.63\%) \\
        \hspace{3mm}Floxif & 1  (0.13\%) \\
        \hspace{3mm}Gogo & 10  (1.25\%) \\
        \hspace{3mm}Morefi & 5  (0.63\%) \\
        \hspace{3mm}Neshta & 5  (0.63\%) \\
        \hspace{3mm}Nuqel & 1  (0.13\%) \\
        \hspace{3mm}Pidgeon & 36  (4.51\%) \\
        \hspace{3mm}Ramnit & 1  (0.13\%) \\
        \hspace{3mm}Sality & 1  (0.13\%) \\
        \hspace{3mm}Sfone & 22  (2.76\%) \\
        \hspace{3mm}Shodi & 11  (1.38\%) \\
        \hspace{3mm}Sivis & 4  (0.50\%) \\
        \hspace{3mm}Small & 2  (0.25\%) \\
        \hspace{3mm}Viking & 1  (0.13\%) \\
        \hspace{3mm}Virut & 1  (0.13\%) \\
        \hspace{3mm}Vobfus & 9  (1.13\%) \\

    \midrule
        \textbf{Total}    & \textbf{ 798}      \\
       \bottomrule
 \end{tabular}}
\end{table}

\begin{table*}[!htp]
\caption{Top dropped extensions in the two experiments. A malware sample tends to drop more files in \thesystem when compared to a virtualized environment.}
    \label{tab:dropped_extensions}
\small
\center{\sffamily
\resizebox{\textwidth}{!}{
\begin{tabular}{lllll}
      \toprule
        \textbf{Extension} & \textbf{\thesystem} &  \textbf{Virtualized Environment} &  \textbf{Difference (\# | \%)}  &  \textbf{Description}  \\
    \midrule
     \textbf{Archive}  &\textbf{12,309}  &\textbf{9,195}  & ~  & ~\\
        \hspace{3mm}cab & 3,504 & 1,989 & 1,515 | 76.2\% & Microsoft cabinet file - compressed archive \\
        \hspace{3mm}pak & 8,805 & 7,206 & 1,599 | 22.2\% & Game Archive/Skype Language Pack \\
    \textbf{Binary} &\textbf{3,132,3425}  &\textbf{1,402,507}  & ~  & ~\\
        \hspace{3mm}api & 22,594 & 7,480 & 15,114 | 202.1\% & Adobe Acrobat Plug-in/WebObjects API File \\
        \hspace{3mm}appx & 2,553 & 557 & 1,996 | 358.3\% & Microsoft Windows 8 app package \\
        \hspace{3mm}bin & 4,891 & 2,053 & 2,838 | 138.2\% & Binary archive \\
        \hspace{3mm}cur & 6,772 & 2,070 & 4,702 | 227.1\% & Windows custom cursor \\
        \hspace{3mm}dat & 5,600 & 3,156 & 2,444 | 77.4\% & General data file \\
        \hspace{3mm}dll & 333,946 & 71,361 & 262,585 | 368.0\% & Windows Dynamic Linked Library\\
        \hspace{3mm}exe & 2,160,744 & 1,114,617 & 1,046,127 | 93.9\% & Executable file \\
        \hspace{3mm}js & 586,494 & 197,952 & 388,542 | 196.3\% & JavaScript file \\ 
        \hspace{3mm}pmp & 2,674 & 917 & 1,757 | 191.6\% & AutoCAD plot model parameter file \\
        \hspace{3mm}pyd & 1,278 & 789 & 489 | 62.0\% & Windows binary that contains compiled Python code \\
        \hspace{3mm}sequ & 4,879 & 1,555 & 3,324 | 213.8\% & Adobe Acrobat Batch sequence \\

    \textbf{Document} &\textbf{74,340}  &\textbf{34,706}  & ~  & ~\\
       \hspace{3mm} html & 32,586 & 13380 & 19206 | 143.5\% & Hypertext Markup Language Document \\ 
       \hspace{3mm} mpp & 2,888 & 924 & 1,964 | 212.6\% & Microsoft Project document \\
       \hspace{3mm} pdf & 19,595 & 13,321 & 6,274 | 47.1\% & Portable Document Format Document \\
       \hspace{3mm} rtf & 1,355 & 910 & 445 | 48.9\% & Rich Text Format document \\
       \hspace{3mm} x3d & 4,662 & 1,391 & 3,271 | 235.2\% & X3D (XML) scene to represent 3D graphics \\
       \hspace{3mm} xml & 13,254 & 4,780 & 8,474 | 177.3\% & Extensible Markup Language File \\

    \textbf{Font} &\textbf{30,719}  &\textbf{10,164}  & ~  & ~\\
        \hspace{3mm} eot & 2,697 & 871 & 1,826 | 209.6\% & Microsoft Embedded OpenType font \\ 
       \hspace{3mm} otf & 25,360 & 7,586 & 17,774 | 234.3\% & Font file \\
       \hspace{3mm} woff & 2,662 & 1,707 & 955 | 55.9\% & Web Open Font Format \\

    \textbf{Image} &\textbf{735,210}  &\textbf{306,284}  & ~  & ~\\
        \hspace{3mm} bmp & 1,383 & 169 & 1,214 | 718.3\% & Bitmap image \\
        \hspace{3mm} gif & 75,298 & 24,718 & 50,580 | 204.6\% & Bitmap image \\
        \hspace{3mm} ico & 28,806 & 18,479 & 10,327 | 55.9\% & Icon file \\
        \hspace{3mm} jpg & 16,865 & 5,462 & 11,403 | 208.8\% & JPEG Bitmap image \\
        \hspace{3mm} png & 266,327 & 136,027 & 130,300 | 95.8\% & Portable Network Graphics image \\
        \hspace{3mm} svg & 346,531 & 121,429 & 225,102 | 185.4\% & Scalable Vector Graphic \\

    \textbf{Raw} &\textbf{328,714}  &\textbf{106,443}  & ~  & ~\\
        \hspace{3mm} aapp & 35,181 & 10,773 & 24,408 | 226.6\% & Adobe Acrobat AcroApp script \\ 
        \hspace{3mm} css & 46,271 & 15,218 & 31,053 | 204.1\% & Cascading style sheets \\ 
        \hspace{3mm} dic & 1,873 & 627 & 1,246 | 198.7\% & Microsoft Office custom dictionary \\
        \hspace{3mm} ini & 13,668 & 4,937 & 8,731 | 176.8\% & Initialization file \\
        \hspace{3mm} json & 21,681 & 1,430 & 20,251 | 1416.2\% & JavaScript Object Notation \\
        \hspace{3mm} lnk & 1,160 & 912 & 248 | 27.2\% & Windows shortcut \\
        \hspace{3mm} log & 15,285 & 7,987 & 7,298 | 91.4\% & General log \\
        \hspace{3mm} tmp & 84,564 & 34,789 & 49,775 | 143.1\% & General temporary file \\
        \hspace{3mm} txt & 109,031 & 29,770 & 79,261 | 266.2\% & Plain text document \\

    \textbf{Audio} &\textbf{2,924}  &\textbf{2,843}  & ~  & ~\\
        \hspace{3mm} wav & 2,924 & 81 & 2,843 | 3509.9\% & Windows waveform sound \\    
        
       \bottomrule
 \end{tabular}}
 }
\end{table*}

\begin{sidewaystable*}[ht]
\caption{I/O Request Packets (IRPs) Types supported by \thesystem}
\scriptsize
\center{\sffamily
\resizebox{\textwidth}{!}{
\begin{tabular}{llll}
      \toprule
        \textbf{Major} & \textbf{Description} & \textbf{Minor} & \textbf{Description} \\
    \midrule
    \textbf{Standard} & ~ & IRP\_MN\_REGINFO & Driver registry path \\ \cline{1-2}
     IRP\_MJ\_CREATE & Open object handle & IRP\_MN\_QUERY\_DIRECTORY & Get Files in directory\\
    IRP\_MJ\_CREATE\_NAMED\_PIPE & Create or open named pipe & IRP\_MN\_NOTIFY\_CHANGE\_DIRECTORY & Notify directory changes\\
    IRP\_MJ\_CLOSE & Close open handle &IRP\_MN\_USER\_FS\_REQUEST & Filesystem requests\\
    IRP\_MJ\_READ & Data read &IRP\_MN\_MOUNT\_VOLUME & Request volume mount\\
    IRP\_MJ\_WRITE & Data write &IRP\_MN\_VERIFY\_VOLUME & Mounted volume integrity check\\
    IRP\_MJ\_QUERY\_INFORMATION & Query object information &IRP\_MN\_LOAD\_FILE\_SYSTEM & Load driver from filesystem\\
    IRP\_MJ\_SET\_INFORMATION & Modify object information &IRP\_MN\_TRACK\_LINK & Watch for device events\\
    IRP\_MJ\_QUERY\_EA & Query object extended attributes &IRP\_MN\_LOCK & Acquire file lock\\
    IRP\_MJ\_SET\_EA & Set object extended attributes &IRP\_MN\_UNLOCK\_SINGLE & Release file lock\\
    IRP\_MJ\_FLUSH\_BUFFERS & Flush cached data to disk &IRP\_MN\_UNLOCK\_ALL & Release all file locks\\
    IRP\_MJ\_QUERY\_VOLUME\_INFORMATION & Query volume information &IRP\_MN\_UNLOCK\_ALL\_BY\_KEY & Release locks associated with key\\
    IRP\_MJ\_SET\_VOLUME\_INFORMATION & Modify volume property &IRP\_MN\_NORMAL & Perform device management tasks via messages\\
    IRP\_MJ\_DIRECTORY\_CONTROL & Directory management operations &IRP\_MN\_DPC & Deferred procedure calls (DPC)\\
    IRP\_MJ\_FILE\_SYSTEM\_CONTROL & Advance file system operations &IRP\_MN\_MDL & Manage memory descriptors lists (MDL)\\
    IRP\_MJ\_DEVICE\_CONTROL & Use device driver functions &IRP\_MN\_COMPLETE & Mark filesystem operation complete\\
    IRP\_MJ\_INTERNAL\_DEVICE\_CONTROL & Midware between device and kernel &IRP\_MN\_COMPRESSED & Read and transfer compressed data\\
    IRP\_MJ\_SHUTDOWN & System shutdown &IRP\_MN\_MDL\_DPC & Merge DPC with completion alert\\
    IRP\_MJ\_LOCK\_CONTROL & Managing byte-range locks & IRP\_MN\_QUERY\_ALL\_DATA & Retrieve comprehensive device information\\
    IRP\_MJ\_CLEANUP & Release process file resources &IRP\_MN\_COMPLETE\_MDL\_DPC & Merge DPC, MDL with completion alert\\
    IRP\_MJ\_CREATE\_MAILSLOT & Manage mailslots for communication &IRP\_MN\_SCSI\_CLASS & Manage SCSI devices\\
    IRP\_MJ\_QUERY\_SECURITY & Query security info about objects &IRP\_MN\_START\_DEVICE & Starting and initializing devices\\
    IRP\_MJ\_SET\_SECURITY & Set object security information &IRP\_MN\_QUERY\_REMOVE\_DEVICE & Pending device removal\\
    IRP\_MJ\_POWER & Manage device power state &IRP\_MN\_REMOVE\_DEVICE & Pending device removal tasks\\
    IRP\_MJ\_SYSTEM\_CONTROL & Request system-wide operations from OS &IRP\_MN\_CANCEL\_REMOVE\_DEVICE & Cancel device removal\\
    IRP\_MJ\_DEVICE\_CHANGE & Manage device events &IRP\_MN\_STOP\_DEVICE & Stop device\\
    IRP\_MJ\_QUERY\_QUOTA & Volume quota information &IRP\_MN\_QUERY\_STOP\_DEVICE & Query stop device possibility\\
    IRP\_MJ\_SET\_QUOTA & Set volume quota &IRP\_MN\_CANCEL\_STOP\_DEVICE & Cancel stop device\\
    IRP\_MJ\_PNP & Manage Plug-and-play devices &IRP\_MN\_QUERY\_DEVICE\_RELATIONS & Discover device relationships\\
    IRP\_MJ\_TRANSACTION\_NOTIFY & Manage operations notifications &IRP\_MN\_QUERY\_INTERFACE & Supported device interfaces\\
    \cline{1-2}
    \textbf{Fast IO} & ~ & IRP\_MN\_SET\_LOCK & Manage object lock requests \\ \cline{1-2}
    IRP\_MJ\_FAST\_IO\_CHECK\_IF\_POSSIBLE & Check object fast I/O capability & IRP\_MN\_QUERY\_CAPABILITIES & Query device capabilities\\
    IRP\_MJ\_DETACH\_DEVICE  & Detach device and free resources &IRP\_MN\_QUERY\_RESOURCES & Query device resource requirements \\
    IRP\_MJ\_NETWORK\_QUERY\_OPEN & Query network connection information &IRP\_MN\_QUERY\_RESOURCE\_REQUIREMENTS & Query device resource requirements\\
    IRP\_MJ\_MDL\_READ   & Read from file or device via MDL & IRP\_MN\_QUERY\_DEVICE\_TEXT & Query device description and location\\
    IRP\_MJ\_MDL\_READ\_COMPLETE & Notify MDL object read completion &IRP\_MN\_FILTER\_RESOURCE\_REQUIREMENTS & Modify device resource requirement\\
    IRP\_MJ\_PREPARE\_MDL\_WRITE & Prepare file for MDL write &IRP\_MN\_READ\_CONFIG & Read connected device configuration\\
    IRP\_MJ\_MDL\_WRITE\_COMPLETE & Complete and notify of MDL write &IRP\_MN\_WRITE\_CONFIG & Write connected device configuration\\
    IRP\_MJ\_VOLUME\_MOUNT  & Mount a volume &IRP\_MN\_EJECT & Remove device from system\\
    IRP\_MJ\_VOLUME\_DISMOUNT & Dismount a volume & IRP\_MN\_DISABLE\_COLLECTION & Prevent automatic resource collection for device\\
    \cline{1-2}
    \textbf{FsFilter} & ~ & IRP\_MN\_EXECUTE\_METHOD & Invoke device functions via its driver \\ \cline{1-2}
    IRP\_MJ\_ACQUIRE\_FOR\_SECTION\_SYNCHRONIZATION & Acquire memory block for synchronization &IRP\_MN\_QUERY\_ID & Uncover the identity of a device\\
    IRP\_MJ\_RELEASE\_FOR\_SECTION\_SYNCHRONIZATION & Release sync memory block &IRP\_MN\_QUERY\_PNP\_DEVICE\_STATE & Query PnP device state\\
    IRP\_MJ\_ACQUIRE\_FOR\_MOD\_WRITE  & Acquire memory section for modifier write &IRP\_MN\_QUERY\_BUS\_INFORMATION & Query connected bus information\\
    IRP\_MJ\_RELEASE\_FOR\_MOD\_WRITE & Release modifier write memory &IRP\_MN\_DEVICE\_USAGE\_NOTIFICATION & Notify of device power usage\\
    IRP\_MJ\_ACQUIRE\_FOR\_CC\_FLUSH & Acquire memory for cache coherency flush &IRP\_MN\_SURPRISE\_REMOVAL & Notify of unexpected device removal\\
    IRP\_MJ\_RELEASE\_FOR\_CC\_FLUSH & Release memory for cache coherency flush &IRP\_MN\_QUERY\_LEGACY\_BUS\_INFORMATION & Query legacy bus hardware information\\
    IRP\_MJ\_NOTIFY\_STREAM\_FO\_CREATION & Notify driver of file create & IRP\_MN\_WAIT\_WAKE & Enable wake from low-power state\\
    ~ & ~ &IRP\_MN\_POWER\_SEQUENCE & Manage device power during dransitions\\
    ~ & ~ &IRP\_MN\_SET\_POWER & Change power state of device\\
    ~ & ~ &IRP\_MN\_QUERY\_POWER & Query ability to move to a named power state\\
    ~ & ~ &IRP\_MN\_QUERY\_SINGLE\_INSTANCE & Get info about named instance of device\\
    ~ & ~ &IRP\_MN\_CHANGE\_SINGLE\_INSTANCE & Modify named instance of device\\
    ~ & ~ &IRP\_MN\_CHANGE\_SINGLE\_ITEM & Modify named device property\\
    ~ & ~ &IRP\_MN\_ENABLE\_EVENTS & Enable specific event generation from device\\
    ~ & ~ &IRP\_MN\_DISABLE\_EVENTS & Disable device from events generation\\
    ~ & ~ &IRP\_MN\_ENABLE\_COLLECTION & Enable automatic resource collection for device\\    
    \bottomrule
 \end{tabular}}}
  \label{tab:irp descriptions}
\end{sidewaystable*}

\end{document}